\documentclass[a4paper,12pt]{article}
\pdfoutput=1
\usepackage{amssymb}
\usepackage{amsmath}
\usepackage{epsfig}
\usepackage{graphicx}

\makeatletter

\@addtoreset{equation}{section}
\makeatother
\def\be{\begin{equation}}
\def\ee{\end{equation}}
\def\bea{\begin{eqnarray}}
\def\eea{\end{eqnarray}}

\def\({\left(}
\def\){\right)}
\def\<{\left<}
\def\>{\right>}

\def\be{\begin{equation}}
\def\ee{\end{equation}}
\def\bea{\begin{eqnarray*}}
\def\eea{\end{eqnarray*}}
\def\ben{\begin{eqnarray}}
\def\een{\end{eqnarray}}
\def\({\left(}
\def\){\right)}
\def\<{\left<}
\def\>{\right>}
\def\!{\right|}
\def\|{\left|}

\def\[{\left[}
\def\]{\right]}

\def\+{\bar}
\def\mb{\mathbb}

\def\D{{\cal{D}}}
\def\L{{\cal{L}}}
\def\t{\widetilde}
\def\A{{\cal{A}}}

\def\F{{\mathcal{F}}}
\def\G{{\mathcal{G}}}

\def\W{{\cal{W}}}

\def\L{{\cal{L}}}

\def\eps{{\cal{\varepsilon}}}

\def\E{{\cal{E}}}

\def\F{{\cal{F}}}

\def\h{\widehat}

\begin{document}

\setlength{\unitlength}{1mm}

\pagestyle{empty}
\vskip-10pt
\vskip-10pt
\hfill 
\begin{center}
\vskip 3truecm
{\Large \bf
Dimensional reduction of M5 branes}
\vskip 2truecm
{\large \bf
Andreas Gustavsson}
\vspace{1cm} 
\begin{center} 
Physics Department, University of Seoul, Seoul 02504 KOREA
\end{center}
\vskip 0.7truecm
\begin{center}
(\tt agbrev@gmail.com)
\end{center}
\end{center}
\vskip 2truecm
{\abstract We study dimensional reduction of M5 branes on a circle bundle when the supersymmetry parameter is not constant along the circle. When the gauge group is Abelian and the fields appear quadratically in the Lagrangian, we can always obtain a supersymmetric five-dimensional theory by keeping fermionic nonzero modes that match with the corresponding nonzero modes of the supersymmetry parameter, and by keeping the zero modes for the bosonic fields as usual. But a supersymmetric non-Abelian generalization can be found only under special circumstances. One instance where we find a non-Abelian supersymmetric generalization is when we perform dimensional reduction along a null direction.}

\vfill
\vskip4pt
\eject
\pagestyle{plain}

\section{Introduction}\label{Introduction}
There is a supersymmetric $(2,0)$ Abelian tensor multiplet in $\mb{R}^{1,5}$ which has a selfdual three-form, five scalar fields and four real Weyl fermionic fields. We can put this tensor multiplet on any six-manifold for which there exists a nontrivial solution to the six-dimensional conformal Killing spinor equation 
\ben
\nabla_M \eps &=& \Gamma_M \eta \label{KE}
\een
Here $M = 0,1,2,3,4,5$ is a vector index on the six-manifold that we will take to be Lorentzian, and $\eps$ will then be the supersymmetry parameter. The equation (\ref{KE}) can be relaxed by turning on supergravity background fields. But we will not study such a generalization here. So $\nabla_M$ here is denoting a curvature covariant derivative that only involves the spin connection and no R-gauge field is turned on.

The classical non-Abelian tensor multiplet is not known and perhaps it does not exist. One approach is then to consider the Abelian tensor multiplet on a circle bundle and perform dimensional reduction along the circle. Then one finds an Abelian 5d Yang-Mills theory for which one can find a non-Abelian generalization. If the supersymmetry parameter is constant along the circle, then it will survive as a supersymmetry under dimensional reduction. Otherwise the supersymmetry will be broken but one may then get a supersymmetric theory by turning on a background R-gauge field that will relax the requirement (\ref{KE}). But that changes the problem that one may want to study. So we would like to analyze whether one can avoid turning on the R-gauge field and somehow take advantage of the fact that the 6d theory is supersymmetric.

One example that one may want to study is the M5 brane on $S^6$ that one may conformally map to $S^1\times H_5$. If one wants to study this problem without any background fields turned on, then one finds that the supersymmetry parameter will have a nontrivial dependence on $S^1$ in $S^1\times H_5$, and dimensional reduction down to $H_5$ yields a non-supersymmetric Yang-Mills theory that is quite difficult to study. Being a nonrenormalizable theory it has no clear well-defined perturbative expansion and there are not many tools to study this theory and supersymmetric localization can not be used if the Yang-Mills theory one gets on $H_5$ is not supersymmetric.

In this paper we will study the following situation. We assume that the 6d theory is supersymmetric on a circle bundle with fiber coordinate $u$. We also assume that the supersymmetry parameter is not constant along $u$. So under dimensional reduction along $u$ all supersymmetry is gone. That is the case if we consider the bosonic and fermionic zero modes. But what if we consider the bosonic zero modes and some fermionic nonzero modes? Is there a consistent trunctation of supersymmetry where bosonic zero modes are kept such that supersymmetry exists in that 5d truncation? 

If the fields appear only quadratically in the Lagrangian so that the gauge group is Abelian, then there always exists such a consistent truncation. To see this, let us schematically write the 6d Lagrangian as
\bea
\L_{6d} &=& (\partial \phi)^2 + \psi \partial \psi
\eea
where $\phi$ denotes bosonic fields and $\psi$ denotes fermionic fields. The supersymmetry variation is schematically on the form
\bea
\delta \phi &=& \eps \psi\cr
\delta \psi &=& \eps \partial \phi
\eea
Then the supersymmetry variation of the Lagrangian is a sum of terms on the form
\bea
0 = \delta \L_{6d} = \sum \partial^2 \phi \eps \psi 
\eea
and the sum vanishes since the 6d Lagrangian is supersymmetric. Now let us make the truncation where we keep the bosonic zero mode along the $u$ direction,
\bea
\phi_0 &=& \int du \phi
\eea
Its supersymmetry variation is 
\ben
\delta \phi_0 &=& \int du \eps \psi\label{uint}
\een
Now let us assume that the supersymmetry parameter has only two nonzero modes,
\bea
\eps &=& e^{i a u} \eps_{+1} + e^{-i a u} \eps_{-1}
\eea
for some real parameter $a$ that depends on the geometry of the six-manifold. This is the generic structure for any solution of (\ref{KE}) on a circle bundle. Here subscripts denote the mode number. Then the integral in (\ref{uint}) picks up corresponding nonzero modes from $\psi$, 
\bea
\delta \phi_0 &=& \eps_{+1} \psi_{-1} + \eps_{-1} \psi_{+1}
\eea
whose supersymmetry variations are
\bea
\delta \psi_{\pm 1} &=& \eps_{\pm 1} \partial \phi_0
\eea
Now let check if the truncated Lagrangian
\bea
\L_{5d} &=& (\partial \phi_0)^2 + \psi_{+1} \partial \psi_{-1} + \psi_{-1} \partial \psi_{+1}
\eea
is supersymmetric. We get
\bea
\delta \L_{5d} &=& \sum \partial^2 \phi_0 \(\eps_{+1} \psi_{-1} + \eps_{-1} \psi_{+1}\)
\eea
but this we can also write as
\bea
\delta \L_{5d} &=& \int du \sum \partial^2 \phi_0 \eps \psi
\eea
Now let us go back to the 6d Lagrangian. If we expand $\phi$ in its Fourier modes as $\phi = \sum_n \phi_n e^{inu}$, then we get
\bea
0 = \delta \L_{6d} = \sum \sum_n \partial^2 \phi_n e^{i n u} \eps \psi
\eea
and we know that this is zero since the 6d Lagrangian is supersymmetric. Of course, if we integrate zero along the fiber, it is still zero, so we have
\bea
0 = \int du \delta \L_{6d} = \sum_n \sum \partial^2 \phi_n \int du e^{i n u} \eps \psi
\eea
If we then put $\phi_n = 0$ for all $n$ except for the zero mode $\phi_0$, then this reduces to 
\bea
0 = \sum \partial^2 \phi_0 \int du \eps \psi = \delta \L_{5d}
\eea
which means that the truncated Lagrangian where only $\phi_0$ is kept, is supersymmetric under the truncated supersymmetries. 

This general argument fails for the non-Abelian generalization where the Lagrangian has higher order terms. For instance if the 6d Lagrangian contains a cubic interaction term of the form $\phi_{+2} \psi_{-1} \psi_{-1}$ and if we have a supersymmetry variation of the form $\delta \phi_{+2} = \eps_{+1} \psi_{+1}$, then the variation of that term will contain a term of the form $\eps_{+1} \psi_{+1} \psi_{-1} \psi_{-1}$ that should survive if the truncation down to the modes $\phi_0$ and $\psi_{\pm 1}$ were a consistent truncation. But we will never get that term if we first truncate the Lagrangian to the modes $\phi_0$ and $\psi_{\pm 1}$ and then make the supersymmetry variation since then we will put the term $\phi_{+2} \psi_{-1} \psi_{-1}$ to zero in that truncated Lagrangian. So the truncation becomes inconsistent in general, when there are higher order terms. However, there can be exceptions where a truncated non-Abelian generalization can be found that is supersymmetric. 

This argument also shows that the critical term to analyze in the supersymmetry variation of the non-Abelian Lagrangian will be the terms that are cubic in the fermionic fields. Typically these term are the most difficult ones to analyse since it usually requires a Fierz rearrangement to see whether the sum of these cubic terms is zero or not. But it is really important to analyze precisely these cubic terms to see whether the non-Abelian Lagrangian is supersymmetric or not. This will become more clear as we proceed with our concrete examples.

In this paper we will study the M5 brane on $\mb{R} \times S^5$ where we have the Lorentzian time along $\mb{R}$. The supersymmetry parameter depends nontrivially on the time direction. First, in section \ref{M5}, we perform dimensional reduction along the time direction and obtain a supersymmetric Abelian Lagrangian. We then show that no non-Abelian generalization exists if we insist on keeping all the supersymmetries of the Abelian theory. In section \ref{The} we reduce the amount of supersymmetry and consider the smaller tensor multiplet that has just one real scalar field. Here we almost seem to find a supersymmetric non-Abelian Lagrangian in 5d by using our truncation, but it turns out to fail. While most terms cancel out nicely, there are cubic terms in the fermionic fields that arises upon a supersymmetry variation and these have to vanish by using a Fierz rearrangement, but these terms do not vanish in that way. We then make a further Weyl projection that reduces supersymmetry further, and then finally we are able to find a supersymmetric Lagrangian. But then, in section \ref{A dual}, we discover that if we make a simple field redefinition, our Lagrangian becomes identical with the Lagrangian that was already found in the literature on $S^5$ \cite{Hosomichi:2012ek} and that was derived from the M5 brane in \cite{Kim:2012ava} by turning on an R-gauge field along the time direction. 

We next consider our second example, in section \ref{Null}, where we consider a null reduction by following closely \cite{Lambert:2020scy}. We take our null direction as a combination of the Hopf circle on $S^5$ and the time direction. We first obtain the Abelian truncated theory and show that it is supersymmetric. We next show that the Abelian theory does not immediately generalize to the non-Abelian case, but if we impose further Weyl projections, then we are able to obtain a non-Abelian Lagrangian. 

There are five appendices. In particular, in appendix \ref{A} we review a 6d formulation of non-Abelian 5d SYM where one introduces an auxiliary geometrical vector field \cite{Lambert:2010wm}, \cite{Gustavsson:2018rcc}, \cite{Gustavsson:2020ugb} and present the closure relations that one gets for these supersymmetry variations and it was this analysis that originally led us to consider the two examples that we are presenting in this paper. Namely these two examples are following from making the two Weyl projections in equations (\ref{W1}) and (\ref{W2}) respectively. The first Weyl projection leads us to the time reduction and the small tensor multiplet. The second Weyl projection leads us to the null reduction.

\section{M5 brane on $\mb{R} \times S^5$}\label{M5}
The six-manifold $\mb{R}\times S^5$ can be conformally mapped to $S^6$ if we assume an Euclidean signature. But here we will assume a Lorentzian signature with time along the $\mb{R}$ direction. Our first goal is to see whether we can derive a supersymmetric theory on $S^5$ from an M5 brane on $\mb{R} \times S^5$ without turning on an R-gauge field along the time direction. The Abelian M5 brane on $\mb{R} \times S^5$ is well-understood. In fact one can generalize to any six-manifold for which (\ref{KE}) has at least one solution. In that case we have the following supersymmetry variations 
\bea
\delta \phi^A &=& i \bar\eps \Gamma^A \psi\cr
\delta B_{MN} &=& i \bar\eps \Gamma_{MN} \psi\cr
\delta \psi &=& \frac{1}{12} \Gamma^{MNP} \eps H_{MNP} + \Gamma^M \Gamma^A \eps \nabla_M \phi^A - 4 \Gamma^A \eta \phi^A
\eea
and the supersymmetric Lagrangian may be expressed as
\bea
\L &=& \L_B - \frac{1}{2} (\nabla_M \phi^A)^2 + \frac{i}{2} \bar\psi \Gamma^M \nabla_M \psi - \frac{R}{10} (\phi^A)^2
\eea
where $\L_B$ is some Lagrangian for the selfdual tensor field whose precise form will not be very important for us now, since we will shortly reduce this Lagrangian down to five dimensions. Here $R$ is the Ricci curvature scalar on six-manifold. We will now specialize to $\mb{R} \times S^5$ and write the metric as
\bea
ds^2 = g_{MN} dx^M dx^N = - dt^2 + G_{mn} dx^m dx^n
\eea
To reduce down to $S^5$, we will represent the gamma matrices in terms of five-dimensional gamma matrices $\gamma^m$ and $\tau^A$ as follows,
\bea
\Gamma^t &=& i \sigma^2 \otimes 1 \otimes 1\cr
\Gamma^m &=& \sigma^1 \otimes \gamma^m \otimes 1\cr
\Gamma^A &=& \sigma^3 \otimes 1 \otimes \tau^A
\eea
The 6d chirality matrix is
\bea
\Gamma &=& \sigma^3 \otimes 1 \otimes 1
\eea
and $\eps$ and $\psi$ have opposite chiralities
\bea
\Gamma \eps &=& - \eps\cr
\Gamma \psi &=& \psi
\eea
and they are Majorana spinors in eleven dimensions,
\bea
\bar\eps &=& \eps^T C_{11d}\cr
\bar\psi &=& \psi^T C_{11d}
\eea
where the Dirac conjugate is defined as $\bar\psi = \psi^{\dag} \Gamma^t$. We may solve (\ref{KE}) by separating its components as
\ben
\partial_t \eps &=& \Gamma_t \eta\cr
\nabla_m \eps &=& \Gamma_m \eta\label{eqs}
\een
We use the relation 
\bea
\Gamma^{mn} \nabla_m \nabla_n \eps &=& - \frac{R}{4} \eps
\eea
where $R = \frac{20}{r^2}$ is the Ricci scalar on $S^5$ with radius $r$, to find the solution
\bea
\eps &=& e^{\frac{i}{2r} t} \(\begin{matrix}
0\\
\E
\end{matrix}\) + e^{-\frac{i}{2r} t} \(\begin{matrix}
0\\
\F
\end{matrix}\)\label{solution}
\eea
We also get
\bea
\eta &=& \frac{i}{2r} e^{\frac{i}{2r} t} \(\begin{matrix}
\E\\
0
\end{matrix}\) - \frac{i}{2r} e^{-\frac{i}{2r} t} \(\begin{matrix}
\F\\
0
\end{matrix}\)  
\eea
Here 
\bea
\nabla_m \E &=& \frac{i}{2r} \gamma_m \E\cr
\nabla_m \F &=& - \frac{i}{2r} \gamma_m \F
\eea
Perhaps the best way to see that this solves (\ref{KE}) is by simply plugging in this solution into (\ref{eqs}) to see that these equations are both satisfied. Let us now study the Majorana condition more closely. The eleven-dimensional charge conjugation matrix is antisymmetric, 
\bea
C_{11d}^T &=& - C_{11d}
\eea
and we will represent it as
\bea
C_{11d} &=& \eps \otimes C \otimes \t{C}
\eea
where $C$ and $\t{C}$ are antisymmetric charge conjugation matrices in 5d, and $\eps$ is the antisymmetric tensor. At this point, things get clearer when we write out all the spinor indices explicitly though, so let us do that here, 
\bea
(C_{11d})_{a\alpha\dot\alpha,b\beta\dot\beta} &=& \eps_{ab} C_{\alpha\beta} C_{\dot\alpha\dot\beta}
\eea
Then the Majorana condition becomes
\bea
(\psi^{a\alpha\dot\alpha})^* i (\sigma^2)^a{}_b &=& \psi^{a\beta\dot\beta} \eps_{ab} C_{\beta\alpha} C_{\dot\beta\dot\alpha}
\eea
We will define the antisymmetric tensor $\eps_{ab}$ such that
\bea
\eps_{+-} &=& 1
\eea
and then we get
\bea
(\psi^{+\alpha\dot\alpha})^* &=& \psi^{+\beta\dot\beta} C_{\beta\alpha} C_{\dot{\beta}\dot{\alpha}}\cr
(\eps^{-\alpha\dot\alpha})^* &=& \eps^{-\beta\dot\beta} C_{\beta\alpha} C_{\dot\beta\dot\alpha}
\eea
From now on we will drop the 6d chirality indices $\pm$ as they play no significant role in 5d. In 5d we do not really have a Majorana condition for the nonzero modes. What we have instead is a relation between $\E$ and $\F$,
\bea
(\E^{\alpha\dot\alpha})^* &=& C_{\alpha\beta} C_{\dot\alpha\dot\beta} \F^{\beta\dot\beta}\cr
(\F^{\alpha\dot\alpha})^* &=& C_{\alpha\beta} C_{\dot\alpha\dot\beta} \E^{\beta\dot\beta}
\eea
These relations follow easily from using the explicit form of our solution, equation (\ref{solution}). But now we would also like to derive the second condition from the first one by taking the complex conjugate. Taking the complex conjugate of the first equation, we get
\bea
\E^{\alpha\dot\alpha} &=& (C_{\alpha\beta})^* (C_{\dot\alpha\dot\beta})^* (\F^{\beta\dot\beta})^*
\eea
We may now multiply by charge conjugation matrices on both sides to get
\bea
C_{\alpha\beta} C_{\dot\alpha\dot\beta} \E^{\beta\dot\beta} &=& C_{\alpha\beta} C_{\dot\alpha\dot\beta} (C_{\beta\gamma})^* (C_{\dot\beta\dot\gamma})^* (\F^{\gamma\dot\gamma})^*
\eea
We shall require that 
\bea
C_{\alpha\beta} (C_{\beta\gamma})^* &=& - \delta_{\alpha}^{\gamma}
\eea
The reason why we put the minus sign here will become clear later on. We can now introduce the inverse 
\bea
C^{\gamma\beta} &=& (C_{\beta\gamma})^*
\eea
We use $C_{\alpha\beta}$ and $C^{\alpha\beta}$ to lower and rise spinor indices by always acting from the left,
\bea
\psi_{\alpha} &=& C_{\alpha\beta} \psi^{\beta}\cr
\psi^{\alpha} &=& C^{\alpha\beta} \psi_{\beta}
\eea
So we define for example
\bea
(\gamma^m)^{\alpha\beta} &=& C^{\beta\gamma} (\gamma^m)^{\alpha}{}_{\gamma}
\eea
We may now find the following relations
\bea
C^{\alpha}{}_{\beta} = C^{\alpha\gamma} C_{\gamma\beta} = \delta^{\alpha}_{\beta}\cr
C_{\alpha}{}^{\beta} = C^{\beta\gamma} C_{\alpha\gamma} = - \delta^{\alpha}_{\beta}
\eea
We have the Fierz expansion of two anticommuting spinors,
\bea
\psi^{\alpha} \psi^{\beta} &=& A C^{\alpha\beta} + B_m (\gamma^m)^{\alpha\beta} + C_{mn} (\gamma^{mn})^{\alpha\beta}
\eea
It corresponds to the following expansion of the tensor product of two spinor representations
\bea
4 \otimes 4 &=& 1_a \oplus 5_a \oplus 10_s
\eea
The subscripts $a$ and $s$ stand for antisymmetric and symmetric representations, so we must have that $C^{\alpha\beta}$ and $(\gamma^m)^{\alpha\beta}$ are antisymmetric, whereas $(\gamma^{mn})^{\alpha\beta}$ is symmetric in $\alpha$ and $\beta$. Our 5d spinor notations follow closely the reference \cite{Hosomichi:2012ek}.

The time direction in Euclidean $\mb{R} \times S^5$ is noncompact if this shall be related by a conformal map to $S^6$. But in Lorentzian signature that we will consider here, the time direction can be taken to be a compact circle with radius $2 \pi r$. We will refrain from discussing any physical implications of having a compact time direction. From a purely mathematical viewpoint of classical supersymmetric field theory, having a compact time direction simply means that we may expand the fields in Fourier modes in the time direction by assuming that time has a periodicity $t \sim t+2\pi r$. For fermions there is as always a possibility of having either periodic or antiperiodic boundary conditions. Since the supersymmetry parameter depends on time through the exponential factors $e^{\pm \frac{i}{2r} t}$ which is antiperiodic as $t$ goes to $t+2\pi r$, we conclude that fermions shall have antiperiodic boundary conditions if we want to have a supersymmetric theory. The bosonic fields must be periodic and therefore only even modes are kept for the bosonic fields, whereas for the fermionic field only the odd modes are kept. And if only the odd modes are kept, it means that there is no fermionic zero mode present. 

But we do not think that we will be able to find a non-Abelian theory if we keep infinitely many Kaluza-Klein modes, neither do we think this is really the right thing to do when the gauge group is non-Abelian because then we shall have instanton particles that are expected to fill in missing modes when we truncate the modes to a finite number of modes. Now instead of truncating to the fermionic zero modes as one normally does in usual dimensional reduction, we will truncate to the lowest lying odd Fourier modes
\bea
\psi &=& e^{\frac{i}{2r} t} \(\begin{matrix}
\chi\\
0
\end{matrix}\) + e^{- \frac{i}{2r} t} \(\begin{matrix}
\zeta\\
0
\end{matrix}\)
\eea
Then the fermionic field has the same type of expansion as the supersymmetry parameter $\eps$ and there is a chance that this will preserve some supersymmetry. There is no Majorana condition on these modes but instead there is a relation between the two modes, 
\bea
(\chi^{\alpha\dot\alpha})^* &=& C_{\alpha\beta} C_{\dot\alpha\dot\beta} \zeta^{\beta\dot\beta}
\eea
The supersymmetry variations can be derived easily by truncating the supesymmetry variations for the Abelian M5 brane. We get
\bea
\delta \phi^A &=& - i \E^{\dag} \tau^A \chi - i \F^{\dag} \tau^A \zeta\cr
\delta A_m &=& - i \E^{\dag} \gamma_m \chi - i \F^{\dag} \gamma_m \zeta\cr
\delta \chi &=& \frac{1}{2} \gamma^{mn} \E F_{mn} - \gamma^m \tau^A \E \nabla_m \phi^A - \frac{2 i}{r} \tau^A \E \phi^A\cr
\delta \zeta &=& \frac{1}{2} \gamma^{mn} \F F_{mn} - \gamma^m \tau^A \F \nabla_m \phi^A + \frac{2 i}{r} \tau^A \F \phi^A
\eea
The corresponding supersymmetric Lagrangian is given by
\bea
\L &=& \frac{1}{4} F_{mn}^2 - \frac{1}{2} (\nabla_m \phi^A)^2 + \frac{i}{2} \chi^{\dag} \gamma^m \nabla_m \chi + \frac{i}{2} \zeta^{\dag} \gamma^m \nabla_m \zeta\cr
&& - \frac{2}{r^2} (\phi^A)^2 + \frac{1}{4r} \(\chi^{\dag} \chi - \zeta^{\dag} \zeta\) 
\eea
The natural choice is to take $\eps$ to be an anti-commuting parameter. In that case the variations of the bosonic fields become hermitian, and we may write these variations as
\bea
\delta \phi^A &=& - i \E^{\dag} \tau^A \chi + i \chi^{\dag} \tau^A \E\cr
\delta A_m &=& - i \E^{\dag} \gamma_m \chi + i \chi^{\dag} \gamma_m \E\cr
\delta \chi &=& \frac{1}{2} \gamma^{mn} \E F_{mn} - \gamma^m \tau^A \E \partial_m \phi^A - \frac{2 i}{r} \tau^A \E \phi^A
\eea
We may also write the Lagrangian as
\bea
\L &=& \frac{1}{4} F_{mn}^2 - \frac{1}{2} (\nabla_m \phi^A)^2 + i \chi^{\dag} \gamma^m \nabla_m \chi\cr
&& - \frac{2}{r^2} (\phi^A)^2 + \frac{1}{2r} \chi^{\dag} \chi
\eea
One may now easily verify that this Lagrangian is invariant under these supersymmetry variations by just using the Killing spinor equation
\bea
\nabla_m \E &=& \frac{i}{2r} \gamma_m \E
\eea
This result is encouraging because it provides our first example of a dimensionally reduced theory that has supersymmetry although the 6d theory has a supersymmetry parameter that depends nontrivially on the circle along which we reduce. Having a supersymmetric Lagrangian, we may also expect that these supersymmetry variations close on some symmetry variations of the Lagrangian. 

However, we will now see that no non-Abelian generalization of this Abelian Lagrangian can be constructed that is supersymmetric. To show this we will proceed iteratively. First we just replace all the derivatives $\nabla_m$ with gauge covariant derivatives $D_m = \nabla_m - i [A_m,\bullet]$ and assume all fields are in the adjoint representation. Then of course the Lagrangian will not be supersymmetric. We then find correction terms such that we cancel the unwanted terms, but such correction terms will also generate new terms that we also need to cancel by adding furher correction terms. This can be analysed fairly systematically. In the end, we will find a fully corrected Lagrangian and corresponding supersymmetry variations but still that Lagrangian will not be supersymmetric. Because of the apparent uniqueness of each term we find in each iteration step, we consider this to be a no-go proof. 

First, if we just replace $\nabla_m$ with $D_m$ everywhere, then we get the following nonvanishing variation of the Lagrangian,
\bea
\delta \L &=& - \frac{1}{2} \chi^{\dag} \gamma^{mn} \tau^A \E [F_{mn},\phi^A] - \chi^{\dag} \gamma^m \E [\phi^A,D_m\phi^A]
\eea
where we define the gauge covariant derivative so that 
\bea
[D_m,D_n] \phi &=& - i [F_{mn},\phi]
\eea
We next cancel both these terms by adding to the Lagrangian the following coupling term
\bea
\L_1 &=& \chi^{\dag} \tau^A [\chi,\phi^A]
\eea
We can not imagine any other term can do this job. But by adding this term, there will be generated some new terms as well, and so now we get
\bea
\delta \L + \delta \L_1 &=& \frac{1}{2} \chi^{\dag} \tau^{AB} \gamma^m \E D_m \([\phi^A,\phi^B]\) + \frac{2i}{r} \chi^{\dag} \tau^{AB} \E [\phi^A,\phi^B]
\eea
plus some cubic terms in $\chi$ that we will not need to analyse further here. Now these two terms can be canceled by modifying the supersymmetry variation by adding the term
\bea
\delta_1 \chi &=& \frac{i}{2} \tau^{AB} \E [\phi^A,\phi^B]
\eea
to $\delta \chi$. But that will also generate another term
\bea
\delta_1 \L_1 &=& i \chi^{\dag} \tau^C \E [[\phi^A,\phi^C],\phi^C]
\eea
but that we can easily cancel by adding the term
\bea
\L_2 &=& - \frac{1}{4} [\phi^A,\phi^B]^2
\eea
But even when taking into account all these non-Abelian correction terms, we will still end up with a nonvanishing variation
\bea
\(\delta + \delta_1\) \(\L + \L_1\) &=& \(- \frac{5i}{4r} + \frac{i}{4r} + \frac{2i}{r}\) \chi^{\dag} \tau^{AB} \E [\phi^A,\phi^B]\cr
&=& \frac{i}{r} \chi^{\dag} \tau^{AB} \E [\phi^A,\phi^B]
\eea
plus those cubic terms in the fermionic fields that we did not analyse here since it is already clear that no non-Abelian Lagrangian can be found. There now is no further terms that we can add that could cancel this nonvanishing variation. This finishes our no-go proof.

\subsection{The small vector multiplet}\label{The}
We may be more successful with finding a non-Abelian generalization if we make our tensor multplet smaller. To this end we will impose the Weyl projection 
\ben
\tau^5 \E &=& \E\label{R}
\een
on the supersymmetry parameter, thus reducing the amount of supersymmetry by half. This will reduce the R-symmetry as $SO(5) \rightarrow SU(2)_R$. But of course, by selecting the fifth direction in (\ref{R}), we will just break $SO(5) \rightarrow SO(4) = SU(2)_F \times SU(2)_R$ but the $SU(2)_F$ will not rotated the supercharges, it will be a flavor symmetry. The original Abelian tensor multiplet breaks into one smaller tensor multiplet with just one real scalar field $\phi = \phi^5$ and a fermionic field that is also subject to the Weyl projection
\bea
\tau^5 \psi &=& \psi
\eea
Then the remaining fields are four real scalars, and another fermionic field subject to the opposite Weyl projection $\tau^5 \psi = - \psi$. These fields form a hypermultiplet. We will discard this hypermultiplet and only focus on the small tensor multiplet. 

Let us now introduce some index notations for the R-symmery. We denote a spinor as 
\bea
\psi^{\alpha\dot\alpha} &=& \(\begin{matrix}
\psi^{\alpha}_I\\
\psi^{\alpha A}
\end{matrix}\)
\eea
The flavor index $A$ is a two-component spinor index that shall not be confused with the $SO(5)$ vector index $A$. We define the gamma matrices $\tau^A = (\tau^i,\tau^5)$ as
\bea
\tau^i &=& \(\begin{matrix}
0 & \sigma^i_{IB}\cr
\sigma^{i,AJ} & 0
\end{matrix}\)\cr
\tau^5 &=& \(\begin{matrix}
\delta_I^J & 0\\
0 & - \delta^A_B
\end{matrix}\)
\eea
The supersymmetry parameter that satisfies $\tau^5 \E = \E$ has a nonvanishing component $\E_I$,
\bea
\E &=& \(\begin{matrix}
\E_I\\
0
\end{matrix}\)
\eea
The antisymmetric charge conjugation matrix is represented as
\bea
C_{\dot\alpha\dot\beta} &=& \(\begin{array}{cc}
\eps^{IJ} & 0\\
0 & \eps_{AB}
\end{array}\)
\eea
We have 
\ben
(\E_I^{\alpha})^* &=& C_{\alpha\beta} \eps^{IJ} \F_J^{\beta}\cr
(\chi_I^{\alpha})^* &=& C_{\alpha\beta} \eps^{IJ} \zeta_J^{\beta}\label{real}
\een
The Killing spinor equations are 
\bea
\nabla_m \E_I &=& \frac{i}{2r} \gamma_m \E_I\cr
\nabla_m \F_I &=& - \frac{i}{2r} \gamma_m \F_I
\eea
The derivation of the second equation from the first by taking the complex conjugate is as follows,
\bea
\nabla_m \F^{\alpha}_I &=& C^{\alpha\beta} \eps_{IJ} (\nabla_m \E^{\beta}_J)^*\cr
&=& - \frac{i}{2r} C^{\alpha\beta} \eps_{IJ} (\gamma_m)^{\gamma}{}_{\beta} (\E^{\gamma}_J)^*\cr
&=& - \frac{i}{2 r} (\gamma^m)^{\gamma\alpha} \eps_{IJ} (\E^{\gamma}_J)^*\cr
&=& \frac{i}{2r} (\gamma^m)^{\alpha\gamma} \eps_{IJ} (\E^{\gamma}_J)^*\cr
&=& - \frac{i}{2r} (\gamma^m)^{\alpha}{}_{\beta} C^{\beta\gamma} \eps_{IJ} (\E^{\gamma}_J)^*\cr
&=& - \frac{i}{2r} (\gamma^m)^{\alpha}{}_{\beta} \F^{\beta}_I
\eea
The supersymmetry variations for the small tensor multiplet are
\bea
\delta \phi &=& - i (\E_I)^{\dag} \chi_I - i (\F_I)^{\dag} \zeta_I\cr
\delta A_m &=& - i (\E_I)^{\dag} \gamma_m \chi_I - i (\F_I)^{\dag} \gamma_m \zeta_I\cr
\delta \chi_I &=& \frac{1}{2} \gamma^{mn} \E_I F_{mn} - \gamma^m \E_I D_m \phi - \frac{2 i}{r} \E_I \phi
\eea
and the supersymmetric Lagrangian is
\bea
\L &=& \frac{1}{4} F_{mn}^2 - \frac{1}{2} (D_m \phi)^2 - \frac{2}{r^2} \phi^2\cr
&& + i (\chi_I)^{\dag} \gamma^m D_m \chi_I + \frac{1}{2r} (\chi_I)^{\dag} \chi_I
\eea
The closure relations for these supersymmetry variations are highly nonstandard,
\bea
[\delta_2,\delta_1] \phi &=& 2 i \L_v \phi\cr
[\delta_2,\delta_1] A_m &=& 2 i \L_v A_m + D_m \Lambda\cr
[\delta_2, \delta_1] \chi_I &=& 8 i \L_B \chi_I + \frac{12}{r} A_I{}^J \chi_J\cr
&& - i \(3 A_I{}^J + 3 B_{pI}{}^J \gamma^p - C_{pqI}{}^J \gamma^{pq}\) \(\gamma^m \nabla_m \chi_J + \frac{1}{2r} \chi_J\)\cr
&& + 8 i \L_{\t{B}} \chi_I - \frac{16}{r} \t{A}_I{}^J \zeta_J  - \frac{4}{r} \t{B}_{mI}{}^J \gamma^m \zeta_J\cr
&& - i \(3 \t{A}_I{}^J + 3 \t{B}_{pI}{}^J \gamma^p - \t{C}_{pqI}{}^J \gamma^{pq}\) \(\gamma^m \nabla_m \zeta_J - \frac{1}{2r} \zeta_J\)
\eea
where
\bea
\L_B \chi_I &:=& B^m{}_I{}^J \nabla_m \chi_J + \frac{1}{4} \nabla_m B_{nI}{}^J \gamma^{mn} \chi_J = B^m{}_I{}^J \nabla_m \chi_J + \frac{i}{2r} C_{mnI}{}^J \gamma^{mn} \chi_J\cr
\L_{\t{B}} \zeta_I &:=& \t{B}^m{}_I{}^J \nabla_m \zeta_J + \frac{1}{4} \nabla_m \t{B}_{nI}{}^J \gamma^{mn} \zeta_J = \t{B}^m{}_I{}^J \nabla_m \zeta_J
\eea
Here the various coefficients are defined as
\bea
\E_I (\E_J)^{\dag} &=& A^J{}_I + B_m{}^J{}_I \gamma^m + C_{mn}{}^J{}_I \gamma^{mn}\cr
\E_I (\F_J)^{\dag} &=& \t{A}^J{}_I + \t{B}_m{}^J{}_I \gamma^m + \t{C}_{mn}{}^J{}_I \gamma^{mn}
\eea
where
\bea
A^J{}_I &=& - \frac{1}{4} (\E_J)^{\dag} \E_I\cr
B_m{}^J{}_I &=& - \frac{1}{4} (\E_J)^{\dag} \gamma_m \E_I\cr
C_{mn}{}^J{}_I &=& \frac{1}{8} (\E_J)^{\dag} \gamma_{mn} \E_I
\eea
and 
\bea
\t{A}^J{}_I &=& - \frac{1}{4} (\F_J)^{\dag} \E_I\cr
\t{B}_m{}^J{}_I &=& - \frac{1}{4} (\F_J)^{\dag} \gamma_m \E_I\cr
\t{C}_{mn}{}^J{}_I &=& \frac{1}{8} (\F_J)^{\dag} \gamma_{mn} \E_I
\eea
There are the following differential relations between these coefficients that one may derive by using the Killing spinor equations,
\bea
\nabla_m A^J{}_I &=& 0\cr
\nabla_m B_m{}^J{}_I &=& \frac{2 i}{r} C_{mn}{}^J{}_I\cr
\nabla_m \t{A}^J{}_I &=& - \frac{i}{r} \t{B}_m{}^J{}_I\cr
\nabla_m \t{B}_n{}^J{}_I &=& \frac{i}{r} \t{A}^J{}_I G_{mn}
\eea

These closure relations reflect the fact that there are many more fermionic degrees of freedom than there are bosonic ones, so closure on the fermion does not give back the same fermion translated or gauge transformed, but instead it maps us back to into a linear combination of $\chi_I$ and $\zeta_I$. 

Let us now turn our attention to a non-Abelian Lagrangian
\bea
\L &=& \frac{1}{4} F_{mn}^2 - \frac{1}{2} (D_m \phi)^2 - \frac{2}{r^2} \phi^2\cr
&& + i (\chi_I)^{\dag} \gamma^m D_m \chi_I + \frac{1}{2r} (\chi_I)^{\dag} \chi_I + e (\chi_I)^{\dag} [\chi_I,\phi]
\eea
and first examine whether this Lagrangian is supersymmetric. This is indeed straightforward to show for all terms, except for the cubic terms in the fermionic fields,
\bea
T &:=& e (\chi_I)^{\dag} [\chi_I,\delta \phi] + i (\chi_I)^{\dag} \gamma^m (- i e) [\delta A_m,\chi_I]\cr
&=& - i e (\E^{\gamma}_J)^* \[\chi^{\gamma c}_J (\chi^{\beta a}_I)^* - (\gamma_m)^{\gamma}{}_{\delta} \chi^{\delta c}_J (\chi^{\alpha a}_I)^* (\gamma^m)^{\alpha}{}_{\beta}\]  \chi^{\beta b}_I
\eea
We expand 
\bea
\chi^{\alpha a}_I (\chi^{\beta b}_J)^* &=& \delta^{\alpha}_{\beta} A^{ab}{}_I{}^J + (\gamma^m)^{\alpha}{}_{\beta} B^{ab}_m{}_I{}^J + (\gamma^{mn})^{\alpha}{}_{\beta} C^{ab}_{mn}{}_I{}^J
\eea
and then 
\bea
T &=& 4 i e (\E^{\gamma}_J)^* \[\delta^{\gamma}_{\beta} A^{ca}{}_J{}^I - (\gamma^m)^{\gamma}{}_{\beta} B^{ca}_m{}_J{}^I\] \chi^{\beta b}_I
\eea
Here
\bea
A^{ca}{}_J{}^I &=& - \frac{1}{4} (\chi^{\alpha a}_I)^* \chi^{\alpha c}_J\cr
B^{ca}_m{}_J{}^I &=& - \frac{1}{4} (\chi^{\alpha a}_I)^* (\gamma_m)^{\alpha}{}_{\beta} \chi^{\beta c}_J
\eea
So we have
\bea
T &=& i e (\E^{\gamma}_J)^* \chi^{\gamma c}_I (\chi^{\alpha a}_I)^* \chi^{\alpha b}_J - i e (\E^{\gamma}_J)^* (\gamma^m)^{\gamma}{}_{\beta} \chi^{\beta c}_I (\chi^{\delta a}_I)^* (\gamma_m)^{\delta}{}_{\epsilon} \chi^{\epsilon b}_J
\eea
We now see that we got an expression that looks similiar to the expression that we started with, but with some indices $I$ and $J$ permuted and an overall sign changed. The up-shot of this analysis is that we can not deduce that $T = 0$ from this result. Now, if we repeat the same steps again, then one may expect we will get a similar expression with the indices $I$ and $J$ in the right order, possibly with a different overall factor from what originally had? Let us now examine this in detail. We start by putting the above expression in the form
\bea
T &=& i e (\E^{\gamma}_J)^* \[\chi^{\gamma c}_I (\chi^{\beta a}_I)^* - (\gamma_m)^{\gamma}{}_{\delta} \chi^{\delta c}_I (\chi^{\alpha a}_I)^* (\gamma^m)^{\alpha}{}_{\beta}\]  \chi^{\beta b}_J
\eea
Now if we use the Fierz expansion, then we get
\bea
T &=& - 4 i e (\E^{\gamma}_J)^* \[\delta^{\gamma}_{\beta} A^{ca} - (\gamma^m)^{\gamma}{}_{\beta} B^{ca}_m\] \chi^{\beta b}_J
\eea
where 
\bea
A^{ca} &=& - \frac{1}{4} (\chi^{\alpha a}_I)^* \chi^{\alpha c}_I\cr
B^{ca}_m &=& - \frac{1}{4} (\chi^{\alpha a}_I)^* (\gamma_m)^{\alpha}{}_{\beta} \chi^{\beta c}_I
\eea
So we have
\bea
T &=& - i e (\E^{\gamma}_J)^* \chi^{\gamma c}_J (\chi^{\alpha a}_I)^* \chi^{\alpha b}_I + i e (\E^{\gamma}_J)^* (\gamma^m)^{\gamma}{}_{\beta} \chi^{\beta b}_J (\chi^{\delta a}_I)^* (\gamma_m)^{\delta}{}_{\epsilon} \chi^{\epsilon c}_I
\eea
and we got back the same expression as we started with. So these lines were insufficient to show that $T$ is vanishing, and most probably $T$ is not vanishing. It may be difficult to actually prove it, but the argument we have presented seems sufficiently convincing to us. 

So we conclude that there is no non-Abelian supersymmetric Lagrangian with this amount of supersymmetry. We can reduce the amount of supersymmetry so that the R-symmetry is further reduced from $SU(2)_R$ down to $U(1)_R$ by imposing the Weyl condition
\bea
(\sigma^3)_I{}^J \E_J &=& \E_J
\eea
Then there is just one complex supersymmetry parameter $\E = \E_1$. With this projection, one finds that the component $\chi_2$ does not enter the supersymmetry multiplet as its supersymmetry variation becomes zero,
\bea
\delta \chi_2 = 0
\eea
and so we define $\chi := \chi_1$ for which we find the supersymmetry variations
\bea
\delta \phi &=& - i \E^{\dag} \chi - i \F^{\dag} \zeta\cr
\delta A_m &=& - i \E^{\dag} \gamma_m \chi - i \F^{\dag} \gamma_m \zeta\cr
\delta \chi &=& \frac{1}{2} \gamma^{mn} \E F_{mn} - \gamma^m \E D_m \phi - \frac{2 i}{r} \E \phi
\eea
The Lagrangian is\footnote{However, we still have the Lagrangian for $\chi_2$ as well, 
\bea
\L_2 &=& i (\chi_2)^{\dag} \gamma^m D_m \chi_2 + \frac{1}{2r} (\chi_2)^{\dag} \chi_2 + e (\chi_2)^{\dag} [\chi_2,\phi]
\eea
but this Lagrangian is not supersymmetric since the corresponding cubic term $T$ upon a supersymmetry variation will not be vanishing, but it is now consistent with supersymmetry to truncate to $\chi_2 = 0$ since the supersymmetry variation of $\chi_2$ is vanishing. So then we will simply get $\L_2 = 0$ and we retain supersymmetry of $\L_2$ trivially by putting $\chi_2 = 0$ as a truncation that is consistent with supersymmetry.}
\bea
\L &=& \frac{1}{4} F_{mn}^2 - \frac{1}{2} (D_m \phi)^2 - \frac{2}{r^2} \phi^2\cr
&& + i \chi^{\dag} \gamma^m D_m \chi + \frac{1}{2r} \chi^{\dag} \chi + e \chi^{\dag} [\chi,\phi]
\eea
The Killing spinor equation is 
\bea
\nabla_m \E &=& \frac{i}{2r} \gamma_m \E
\eea
Originally we had
\bea
\F^{\alpha}_2 &=& \eps_{21} C^{\alpha\beta} (\E^{\beta}_1)^*\cr
\zeta^{\alpha}_2 &=& \eps_{21} C^{\alpha\beta} (\chi^{\beta}_1)^*
\eea
Now we define $\F^{\alpha} := \F^{\alpha}_2$ and $\zeta^{\alpha} := \zeta^{\alpha}_2$ so with $\eps^{12} = 1$, we get the relations
\bea
\F^{\alpha} &=& C^{\alpha\beta} (\E^{\beta})^*\cr
\zeta^{\alpha} &=& C^{\alpha\beta} (\chi^{\beta})^*
\eea
Let us now again analyze the cubic terms in the fermionic field that arise upon a supersymmetry variation of this Lagrangian. These terms are
\bea
T &:=& e (\chi)^{\dag} [\chi,\delta \phi] + i (\chi)^{\dag} \gamma^m (- i e) [\delta A_m,\chi]\cr
&=& - i e (\E^{\gamma})^* \[\chi^{\gamma c} (\chi^{\beta a})^* - (\gamma_m)^{\gamma}{}_{\delta} \chi^{\delta c} (\chi^{\alpha a})^* (\gamma^m)^{\alpha}{}_{\beta}\]  \chi^{\beta b}
\eea
We expand 
\bea
\chi^{\alpha a} (\chi^{\beta b})^* &=& \delta^{\alpha}_{\beta} A^{ab} + (\gamma^m)^{\alpha}{}_{\beta} B^{ab}_m + (\gamma^{mn})^{\alpha}{}_{\beta} C^{ab}_{mn}
\eea
and then 
\bea
T &=& 4 i e (\E^{\gamma})^* \[\delta^{\gamma}_{\beta} A^{ca} - (\gamma^m)^{\gamma}{}_{\beta} B^{ca}_m\] \chi^{\beta b}
\eea
Here
\bea
A^{ca} &=& - \frac{1}{4} (\chi^{\alpha a})^* \chi^{\alpha c}\cr
B^{ca}_m &=& - \frac{1}{4} (\chi^{\alpha a})^* (\gamma_m)^{\alpha}{}_{\beta} \chi^{\beta c}
\eea
So we have
\bea
T &=& i e (\E^{\gamma})^* \chi^{\gamma c} (\chi^{\alpha a})^* \chi^{\alpha b} - i e (\E^{\gamma})^* (\gamma^m)^{\gamma}{}_{\beta} \chi^{\beta c} (\chi^{\delta a})^* (\gamma_m)^{\delta}{}_{\epsilon} \chi^{\epsilon b}
\eea
We now see that we got back the same expression as the one we started with, but with an overall minus sign, so $T = - T$, which clearly shows that $T = 0$ and the Lagrangian is supersymmetric.

\subsection{A dual description with an R-gauge field}\label{A dual}
By making a few changes of viewpoint we may recover the theory one gets by turing on an R-gauge field and make contact with the results in \cite{Hosomichi:2012ek}. We relabel the spinor field and its complex conjugate field as 
\bea
\chi &=& \psi_1\cr
\zeta &=& \psi_2
\eea
and similarly 
\bea
\E &=& \E_1\cr
\F &=& \E_2
\eea
Then we may state a Majorana condition as
\bea
\psi_I^{\alpha} &=& \eps_{IJ} C^{\alpha\beta} (\psi^{\beta}_J)^*
\eea
that we get from  
\bea
\zeta^{\alpha} &=& C^{\alpha\beta} (\chi^{\beta})^*
\eea
Moreover, the Killing spinor equations for $\chi$ and $\zeta$ can now be grouped together into one Killing spinor equation for the Majorana spinor $\E_I$ 
\bea
\nabla_m \E_I &=& \frac{i}{2r} (\sigma^3)_I{}^J \gamma_m \E_J
\eea
So there is an exact isomorphism between the theory we get by turning on an R-gauge field, and the theory we get in this entirely different way by keeping nonzero modes for the fermionic field and not turning on any R-gauge field. 

In one viewpoint, $\chi$ and $\zeta$ are nonzero Kaluza-Klein modes who receive an extra mass simply by the fact that they are nonzero modes. In the other viewpoint, $\chi$ and $\zeta$ form two components in an $SU(2)_R$ Majorana spinor which is a zero mode spinor upon dimensional reduction with an R-gauge field turned on and the mass of these fermions is induced from that R-gauge field in the six-dimensional theory. Both ways result in the same 5d theory, but the 6d theories seem to be very different.

Once having realized this kind of dual description, we can proceed and use all knowledge that we already have of this 5d theory from say \cite{Hosomichi:2012ek}. We will review that theory below in order to put it in relation to the 6d theory on $\mb{R} \times S^5$. We will focus only on the case of Abelian gauge group for simplicity. The non-Abelian generalization will be straightforward and can be found in \cite{Hosomichi:2012ek}. We begin by turning on an R-symmetry gauge field to preserve supersymmetry for fermionic zero modes. The Killing spinor equation is modified to 
\bea
D_t \E_I &=& \frac{i}{2r} (\sigma^3)_I{}^J \E_J\cr
\nabla_m \E_I &=& \frac{i}{2r} \gamma_m (\sigma^3)_I{}^J \E_I
\eea
We have the Majorana condition
\bea
(\E^{\alpha}_I)^* &=& C_{\alpha\beta} \eps^{IJ} \E^{\beta}_J
\eea
We also have 
\bea
\eta_I &=& \frac{i}{2r} (\sigma^3)_I{}^J \eps_J
\eea
The supersymmetry variations are
\bea
\delta \phi &=& - i (\E_I)^{\dag} \psi_I\cr
\delta A_m &=& - i (\E_I)^{\dag} \gamma_m \psi_I\cr
\delta \psi_I &=& \frac{1}{2} \gamma^{mn} \E_I F_{mn} - \gamma^m \E_I \partial_m \phi - \frac{2 i}{r} (\sigma^3)_I{}^J \E_J \phi
\eea
With a commuting supersymmetry parameter, we have the following closure relations. Closure on $\phi$,
\bea
\delta^2 \phi &=& i (\E_I)^{\dag} \gamma^m \E_I \partial_m \phi
\eea
Closure on $A_m$,
\bea
\delta^2 A_m &=& i (\E_I)^{\dag} \gamma^n \E_I F_{nm} + \partial_m \(- i (\E_I)^{\dag} \E_I \phi\)
\eea
Closure on $\psi_I$,
\bea
\delta^2 \psi_I &=& - 8 i B^m \nabla_m \psi_I+\frac{12A}{r}(\sigma^3)_I{}^J\psi_J\cr
&& - \(3A+3B_p\gamma^p\)\(i\gamma^m\nabla_m\psi_I+\frac{1}{2r}(\sigma^3)_I{}^J\psi_J\) 
\eea
The supersymmetric Lagrangian is $\L = L_B + \L_F^I + \L_F^{II}$ where
\bea
\L_B &=& \frac{1}{4} F_{mn}^2 - \frac{1}{2} (\nabla_m \phi)^2 - \frac{2}{r^2} \phi^2\cr
\L_F^I &=& \frac{i}{2} (\psi_I)^{\dag} \gamma^m \nabla_m \psi_I\cr
\L_F^{II} &=& \frac{1}{4r} (\psi_I)^{\dag} (\sigma^3)_I{}^J \psi_J
\eea
For an anticommuting supersymmetry parameter, we have 
\bea
\delta \phi &=& i (\psi_I)^{\dag} \E_I\cr
\delta A_m &=& i (\psi_I)^{\dag} \gamma_m \E_I
\eea
and then we get
\bea
\delta \L_B &=& - \nabla_m F^{mn} i (\psi_I)^{\dag}\gamma_n \E_I + \nabla^2 \phi i (\psi_I)^{\dag} \E_I - \frac{4}{r^2} \phi i (\psi_I)^{\dag} \E_I\cr
\delta \L_F^I &=& i (\psi_I)^{\dag} \gamma^m \nabla_m \(\frac{1}{2} \gamma^{pq} \E_I F_{pq} - \gamma^p \E_I \partial_p \phi - \frac{2 i}{r} (\sigma^3)_I{}^J \E_J \phi
\)\cr
&=& i (\psi_I)^{\dag} \gamma_q \E_I \nabla_m F^{mq} - i (\psi_I)^{\dag} \E_I \nabla^2 \phi + \frac{2}{r} (\psi^{\dag} \gamma^m (\sigma^3)_I{}^J \E_J \nabla_m \phi\cr
&& + \frac{i}{2} (\psi_I)^{\dag} \gamma^m \gamma^{pq} (\nabla_m \E_I) F_{pq} - i (\psi_I)^{\dag} \gamma^m \gamma^p (\nabla_m \E_I) \nabla_p \phi + \frac{2}{r} (\psi_I)^{\dag} (\sigma^3)_I{}^J (\gamma^m \nabla_m \E_J) \phi\cr
\delta \L_F^{II} &=& \frac{1}{2r} (\psi_I)^{\dag} (\sigma^3)_I{}^J \(\frac{1}{2} \gamma^{mn} \E_J F_{mn} - \gamma^m \E_J \nabla_m \phi - \frac{2 i}{r} (\sigma^3)_J{}^K \E_K \phi\)
\eea
Using 
\bea
\nabla_m \E_I &=& \frac{i}{2r} (\sigma^3)_I{}^J \E_J
\eea
and 
\bea
\gamma^m \gamma^p \gamma_m &=& - 3 \gamma^p\cr
\gamma^m \gamma^{pq} \gamma_m &=& \gamma^{pq}
\eea
we can show that all terms cancel against each other so that $\delta \L = 0$.

We may take the supersymmetry variations off-shell,
\bea
\delta \phi &=& - i (\E_I)^{\dag} \psi_I\cr
\delta A_m &=& - i (\E_I)^{\dag} \gamma_m \psi_I\cr
\delta \psi_I &=& \frac{1}{2} \gamma^{mn} \E_I F_{mn} - \gamma^m \E_I \partial_m \phi - \frac{i}{r} (\sigma^3)_I{}^J \E_J \phi + \E_J D^J{}_I\cr
\delta D^J{}_I &=& 2 (\E_J)^{\dag} \(i \gamma^m \nabla_m \psi_I + \frac{1}{2r} (\sigma^3)_I{}^L \psi_L\) - \frac{1}{r} (\sigma^3)_I{}^J (\E_K)^{\dag} \psi_K\cr
&& - \delta_I^J (\E_K)^{\dag} \(i \gamma^m \nabla_m \psi_K + \frac{1}{2r} (\sigma^3)_K{}^L \psi_L\)
\eea
where the second line in the variation of $D^J{}_I$ removes the trace part, where we notice that $\sigma^3$ is already traceless. The Lagrangian is
\bea
\L &=& \frac{1}{4} F_{mn}^2 - \frac{1}{2} (\nabla_m\phi)^2\cr
&& + \frac{1}{4} D^I{}_J D^J{}_I + \frac{i}{2r} (\sigma^3)_I{}^J D^I{}_J \phi - \frac{5}{2r^2} \phi^2\cr
&& + \frac{i}{2} (\psi_I)^{\dag} \gamma^m \nabla_m \psi_I + \frac{1}{4r} (\psi_I)^{\dag} (\sigma^3)_I{}^J \psi_J
\eea
Integrating out $D^I{}_J$ amounts to putting
\ben
D^I{}_J &=& - \frac{i}{r} (\sigma^3)_I{}^J \phi\label{saddle}
\een
and then the second line in the Lagrangian becomes
\bea
\frac{1}{2r^2} \phi^2 - \frac{5}{2r^2} \phi^2 &=& - \frac{2}{r^2} \phi^2
\eea
which is the right on-shell action, and also the supersymmetry variation becomes
\bea
\delta \psi_I &=& \frac{1}{2} \gamma^{mn} \E_I F_{mn} - \gamma^m \E_I \partial_m \phi - \frac{2 i}{r} (\sigma^3)_I{}^J \E_J \phi\cr
\delta D^J{}_I &=& - \frac{i}{r} (\sigma^3)_I{}^J \delta \phi
\eea
which are the right on-shell variation. The on-shell variation of $D^J{}_I$ corresponds to a variation of the on-shell saddle point equation (\ref{saddle}). 

But this does not explain why we shall make this funny shift away from say the saddle point value zero for $D^I{}_J$. To understand why we shall construct the Lagrangian such that we have the shifted saddle point value (\ref{saddle}), we look at the supersymmetry variation of the fermionic part of the Lagrangian with $\delta \psi_I = \E_J D^J{}_I$. We notice that there is no term that involves a derivative of $D^J{}_I$ as this field is an auxiliary non-dynamical field. Therefore we shall make an integration by parts such that the variation of the fermionic terms becomes
\bea
\delta \L_F &=& (\delta \psi_I)^{\dag} \(i \gamma^m \nabla_m \psi_I + \frac{1}{2r} (\sigma^3)_I{}^K \psi_K\)
\eea
This is opposite the the convention we used before where made integrations by parts so that no derivatives acted on the fermionic field. But here this new convention makes better sense because we do not get derivative of the auxiliary field from the bosonic terms by varying the auxiliary field. Now let us compute this variation with $(\delta \psi_I)^{\dag} = (D^J{}_I)^* (\E_J)^{\dag}$. We then notice that 
\bea
- (D^J{}_I)^* = \eps_{JK} \eps^{IL} D^K{}_L = \eps_{JK} D^{KI} = \eps_{JK} D^{IK} = D^I{}_J
\eea
where the first equality is a consequence of demanding 
\bea
(\delta \psi_I)^{\dag} &=& \delta (\psi_I)^{\dag}
\eea
with $\delta \psi_I = \E_J D^J{}_I$. Here is the computation. First,
\bea
(\delta \psi^{\alpha}_I)^* = (\E^{\alpha}_J D^J{}_I)^* = C_{\alpha\beta} \eps^{JK} \E^{\beta}_K (D^J{}_I)^*
\eea
and second,
\bea
\delta (\psi^{\alpha}_I)^* = C_{\alpha\beta} \eps^{IJ} \delta \psi^{\beta}_J = C_{\alpha\beta} \eps^{IJ} \E^{\beta}_K D^K{}_J
\eea
Then by identifying these two results, we get
\bea
\eps^{JK} (D^J{}_I)^* &=& \eps^{IJ} D^K{}_J
\eea
After these preliminaries, we get
\bea
\delta \L_F &=& - \frac{1}{2r} D^I{}_J (\E_J)^{\dag} (\sigma^3)_I{}^K \psi_K
\eea
We also get
\bea
\delta \L_B &=& \frac{1}{2r} D^I{}_J (\E_J)^{\dag} (\sigma^3)_I{}^K \psi_K
\eea
and so we see that the sum is zero, $\delta \L_F + \delta \L_B = 0$. This shows that the Lagrangian is supersymmetric.

Offshell closure is slightly modified from onshell closure as follows. We have
\bea
\delta^2 \psi_I &=& 8 i B^m \nabla_m \psi_I + \frac{12 A}{r} (\sigma^3 \psi)_I\cr
\delta^2 D^J{}_I &=& i (\E_K)^{\dag} \gamma^m \E_K \nabla_m D^J{}_I - \frac{3}{r} (\E_L)^{\dag} \E_L (\sigma^3)_K{}^J D^K{}_I
\eea
Now these results can be recast in the form
\bea
\delta^2 \psi_I &=& ... - \frac{3}{2r} (\E_L)^{\dag} \E_L (\sigma^3)_I{}^J \psi_J\cr
\delta^2 D^J{}_I &=& ... - \frac{3}{r} (\E_L)^{\dag} \E_L (\sigma^3)_K{}^J D^K{}_I
\eea
and we see that we got an R-symmetry rotation. Of course the scalar field $\phi$ is an R-symmetry singlet so it will not be R-symmetry rotated.

The results we have found here all followed from straightforward computations. But it remains a mystery to us why two different kind of dimensional reductions result in the same 5d Lagrangian. In one instance we did not turn on any R-gauge field but instead we kept the modes $\phi_0$ and $\psi_{\pm 1}$. In the other instance we turn on an R-gauge field and keep the zero modes $\phi_0$ and $\psi_0$. Both ways lead us to the exact same Lagrangian in 5d if we impose the appropriate Weyl projections, but we do not understand why that is so.

\section{Null reduction}\label{Null}
A general null reduction of the M5 brane was studied in \cite{Lambert:2020scy}. Here we will stay with our example of $\mb{R} \times S^5$ with Lorentzian time along $\mb{R}$ for simplicity, although we believe that our results can be generalized to any Lorentzian six-manifold without any new conceptional difficulties, beyond those we will address here. We will perform the dimensional reduction along the null direction that is formed out of the time direction and a circle fiber direction on $S^5$ when viewed as a circle fiber over $\mb{C}P^2$. However, once we specify a circle fiber, there are two null directions, $x^+$ and $x^-$ and we need to make a choice. We will make the choice such that we perform the dimensional reduction along the $x^-$ direction. This choice of null direction is correlated with some chirality choices for the supersymmetry parameter that we wish to make, as we will now explain.

We start by writing the metric on $\mb{R}\times S^5$ as a metric over the base-manifold $\mb{R} \times \mb{C}P^2$. The M5 brane on (a Hopf circle bundle over) $\mb{R}\times \mb{C}P^2$ was first studied in \cite{Kim:2012tr}. We start by writing the 6d metric in the form
\bea
ds^2 &=& r^2 (dy + \kappa_i dx^i)^2 - dt^2 + G_{ij} dx^i dx^j
\eea
where the five coordinates $x^m$ on $S^5$ are separated as $y \sim y + 2\pi$ for the circle fiber, and $x^i$ for the base manifold $\mb{C}P^2$, and $\kappa_i$ is the graviphoton whose nonvanishing curvature components are 
\bea
w_{\h{1}\h{2}} = w_{\h{3}\h{4}} = \frac{2}{r^2}
\eea
where the hats on these indices indicate that they are tangent space indices of $\mb{C}P^2$. Here we use $G_{ij}$ to denote the 4d metric tensor on $\mb{C}P^2$ whose inverse is denoted $G^{ij}$. Further details regarding this Hopf fibration over $\mb{C}P^2$ can be found in appendix \ref{Pope}. 

We then also split the indices in the 5d Killing spinor equation
\bea
\nabla_m \E^{\alpha\dot\alpha} &=& \frac{i}{2r} (\gamma_m)^{\alpha}{}_{\beta} \E^{\beta\dot\alpha}
\eea
on $S^5$ into two equations
\ben
\nabla_y \E &=& \frac{i}{2r}\gamma_y \E\cr
\nabla_i \E &=& \frac{i}{2r}\gamma_i \E\label{5dKSE}
\een
associated to the fiber and the base-manifold respectively (and from now, we suppess the spinor indices). To analyse these equations further, we need expressions for these covariant derivatives in terms of spin connections and we need to express the 5d gamma matrices and in terms of 4d gamma matrices. To this end, we start by writing down expressions for the vielbein
\bea
e^{\h{t}} &=& dt\cr
e^{\h{y}} &=& r \(dy + \kappa_i dx^i\)\cr
e^{\h{i}} &=& E^{\h{i}}{}_j dx^j
\eea
and its inverse
\bea
e_{\h{t}} &=& \partial_t\cr
e_{\h{y}} &=& \frac{1}{r} \partial_y\cr
e_{\h{i}} &=& E^j{}_{\h{i}} \(\partial_j - \kappa_j \partial_y\)
\eea
Using these vielbeins, we may expand the 5d gamma matrices $\gamma_m$ in terms of 4d gamma matrices $\t{\gamma}_i = E^{\h{i}}{}_i \gamma_{\h{i}}$ and $\gamma := \gamma^{\h{1}\h{2}\h{3}\h{4}}$ as follows,
\bea
\gamma_y &=& r \gamma\cr
\gamma_i &=& \t\gamma_i + r \kappa_i \gamma
\eea
and then we use standard circle bundle expressions for the 5d covariant derivative acting on a 5d spinor $\psi$,
\bea
\nabla_y \psi &=& \partial_y \psi - \frac{r^2}{8} w_{ij} \t\gamma^{ij} \psi\cr
\nabla_i \psi &=& \t{\nabla}_i \psi - \frac{r^2}{8} \kappa_i w_{kl} \t\gamma^{kl} \psi + \frac{r}{4} w_{ij} \t\gamma^j \gamma \psi
\eea
where $\t{\nabla}_i$ denotes the covariant derivative with respect to the metric on the 4d base space. We are now ready to express (\ref{5dKSE}) in 4d quantities,
\bea
\partial_y \E - \frac{r^2}{8} w_{ij} \gamma^{ij} \E &=& \frac{i}{2} \gamma \E\cr
\nabla_i \E - \frac{r^2}{8} \kappa_i w_{kl} \gamma^{kl} \E + \frac{r}{4} w_{ij} \gamma^j \gamma \E &=& \frac{i}{2r} \(\gamma_i + r \kappa_i \gamma\) \E
\eea
where now all quantities are 4d quantities, and so we have dropped the tildes for notational simplicity. We may also express the second equation more simply as
\bea
\D_i \E &=& \frac{i}{2r} \gamma_i \E - \frac{r}{4} w_{ij} \gamma^j \gamma \E
\eea
where we have introduced the curly derivative
\bea
\D_i \psi &=& \nabla_i \psi - \kappa_i \partial_y \psi
\eea
But let us first analyze the first equation. Plugging in the explicit form of $w_{ij}$, this equation reads
\bea
\partial_y \E &=& \frac{1}{2} \(\gamma^{\h{1}\h{2}} + \gamma^{\h{3}\h{4}} + i \gamma\) \E
\eea
Of course the spinor $\E^{\alpha\dot\alpha}$ has four different indices $\alpha$. To see the meaning of these various indices more clearly, we will introduce a spin notation $\alpha = (s_1,s_2)$ where the spins $s_1$ and $s_2$ are defined by
\bea
\frac{i}{2} \gamma^{\h{1}\h{2}} \E &=& s_1 \E\cr
\frac{i}{2} \gamma^{\h{3}\h{4}} \E &=& s_2 \E
\eea
Let us first consider the spinor component $(s_1,s_2) = (+,+)$ where $\pm$ represent spins $\pm \frac{1}{2}$. The Killing spinor equations then reduce to 
\bea
\partial_y \E &=& - \frac{3i}{2} \E\cr
\D_i \E &=& 0
\eea
Moving up to 6d, we have the conformal Killing spinor solution
\bea
\eps &=& e^{\frac{i}{2r} t -\frac{3i}{2} y} \E + e^{-\frac{i}{2r} t + \frac{3i}{2} y} \F
\eea
This is the singlet solution. The other cases are $(s_1,s_2) = \{(-,-),(+,-),(-,+)\}$ that form a triplet. For any of these components, the first Killing spinor equation becomes 
\bea
\partial_y \E &=& \frac{i}{2} \E
\eea
and then the 6d solution becomes
\bea
\eps &=& e^{\frac{i}{2r} t + \frac{i}{2} y} \E + e^{- \frac{i}{2r} t - \frac{i}{2} y} \F
\eea
but the Killing spinor equations for $\E$ and $\F$ now become more complicated. We introduce light cone coordinates
\bea
x^{\pm} &=& \frac{1}{\sqrt{2}} (t\pm r y)
\eea
Expressed in these light-cone coordinates, the singlet solution is
\bea
\eps &=& e^{\frac{i}{r\sqrt{2}} \(- x^+ + 2 x^-\)} \E + e^{- \frac{i}{r\sqrt{2}} \(- x^+ + 2 x^-\)} \F
\eea
and the triplet solutions are 
\bea
\eps &=& e^{\frac{i}{r\sqrt{2}} x^+} \E + e^{- \frac{i}{r\sqrt{2}} x^+} \F
\eea
Since these triplet supersymmetry parameters do not depend on $x^-$, the corresponding supersymmetry survives upon dimensional reduction along $x^-$ without any need to turn on an R-gauge field. While this is nice, the price we have to pay is having a more complicated Killing spinor equation. 

We will study the singlet solution instead. This has a simpler Killing spinor equation, and it gives us an opportuntiy to study a situation where the supersymmetry parameter depends nontrivially on the fiber direction along which we dimensionally reduce. But again the question arises, along which direction we shall reduce. Let us start by recalling the 6d Weyl condition $\Gamma \eps = - \eps$ that we will write as
\ben
\Gamma^{t y} \Gamma^{1234} \eps &=& - \eps\label{p1}
\een
As we mentioned in the Introduction, we also want to impose the Weyl projection
\bea
\Gamma_M \eps v^M &=& 0
\eea
where $v^M$ is now to be either one of the lightcone directions, $v^M = \delta^M_{\pm}$. So the above Weyl projection amounts to 
\bea
\Gamma_{\pm} \eps &=& 0
\eea
where 
\bea
\Gamma_{\pm} &=& \frac{1}{\sqrt{2}} \(\Gamma_t \pm \frac{1}{r} \Gamma_y\)
\eea
so we may also express this Weyl projection as 
\ben
\Gamma^{ty} \eps &=& \mp \eps\label{p2}
\een
Now by combining (\ref{p1}) and (\ref{p2}), we get
\bea
\Gamma^{1234} \eps &=& \pm \eps
\eea
The singlet supersymmetry parameter has $\Gamma^{1234} \eps = - \eps$ and therefore we shall take $v^M = \delta^M_-$ and perform the dimensional reduction along the $x^-$ direction. Let us write down the singlet solution again as
\bea
\eps &=& e^{\frac{i\sqrt{2}}{r} x^-} \E + e^{-\frac{i\sqrt{2}}{r} x^-} \F
\eea
Then upon dimensional reduction, we shall expand the fermionic field in the same modes as
\bea
\psi &=& e^{\frac{i\sqrt{2}}{r} x^-} \chi + e^{-\frac{i\sqrt{2}}{r} x^-} \zeta
\eea
Of course we do not know the non-Abelian supersymmetry variations for the M5 brane. The strategy will therefore be to start with the Abelian supersymmetry variations for the M5 brane, and reduce these along the $x^-$ direction by using the above mode expansion for the fermionic field. We will also find a corresponding Abelian Lagrangian that is supersymmetric. These steps are in parallel with what we have already done when we reduced along the time direction, although the reduction along $x^-$ requires a lot more computations. Once we have obtained these Abelian supersymmetries and Lagrangian, the generalization to the non-Abelian case will be examined. We start by replacing derivatives with gauge covariant derivatives and examine the term in the variation of the Lagrangian that is cubic in the fermionic field. But this term is vanishing, not because of some Fierz rearrangment, but simply because, as we will see, the supersymmetry variation of the following combination of gauge fields is vanishing,\footnote{In 6d we also have the gauge fixing condition $A_M v^M = A_- = 0$ that can be seen as a consequence of $A_M = B_{MN} v^N$.}
\bea
\delta \(A_i - \kappa_i A_y\) &=& 0
\eea
and it is precisely this combination that enters in the kinetic term for the fermionic field
\bea
i \chi^{\dag} \gamma^i \D_i \chi
\eea
So when we vary the gauge potential in this term, there will be no cubic term generated. Let us now show this in more detail. Let us start with the 6d supersymmetry variation   
\bea
\delta A_M &=& - i \bar\psi \Gamma_{MN} \eps v^N
\eea
from which we obtain
\bea
\delta A_i &=& - \frac{i r}{\sqrt{2}} \kappa_i \bar\psi \eps\cr
\delta A_+ &=& - i \bar\psi \eps
\eea
Then 
\bea
\D_i \psi &=& \(D_i - \kappa_i D_y\) \psi\cr
&=& \(D_i - \frac{r}{\sqrt{2}} \kappa_i D_+\) \psi
\eea
The important observation is now that 
\bea
\delta \D_i = - i e \delta \(A_i - \frac{r}{\sqrt{2}} \kappa_i A_+\) = 0
\eea
For this computation we have used
\bea
\Gamma_{\pm} &=& \Gamma_{\h{\pm}}\cr
\Gamma_i &=& \t\Gamma_i + \frac{r}{\sqrt{2}} \(\Gamma_+ - \Gamma_-\)
\eea
and then 
\bea
\Gamma_{i\pm} &=& \t\Gamma_i \Gamma_{\pm} + \frac{r}{\sqrt{2}} \kappa_i \Gamma_{+-}
\eea
We have 
\bea
\Gamma_{\pm} &=& \frac{1}{\sqrt{2}} \(\Gamma^t \pm \frac{1}{r} \Gamma_y\)\cr
\Gamma^{\pm} &=& \frac{1}{\sqrt{2}} \(\Gamma^t \pm r \Gamma^y\)
\eea
and then we get
\bea
\Gamma_{+-} &=& \Gamma^{\h{t}\h{y}}
\eea
We impose the Weyl projection
\bea
\Gamma_- \eps &=& 0
\eea
and then we get
\bea
\Gamma_- \Gamma_+ \eps &=& \(\{\Gamma_-,\Gamma_+\} - \Gamma_+ \Gamma_- \)\eps\cr
&=& - 2 \eps
\eea
where we notice the metric is 
\bea
ds^2 &=& - 2 e^{\h{+}} e^{\h{-}} + e^{\h{i}} e^{\h{i}}
\eea
Now having shown that $\delta \D_i = 0$ is, as we will see below, just one crucial step among many other steps towards obtaining a supersymmetry non-Abelian Lagrangian. 

We begin with assuming the gauge group is Abelian and let us first study the supersymmetry variation of the tensor gauge field in 6d,
\bea
\delta H_{MNP} &=& - 3 i \partial_M \(\bar\psi \Gamma_{NP} \eps\) 
\eea
for an anticommuting supersymmetry parameter, for which we have the relation
\bea
\bar\eps \Gamma_{MN} \psi = (\eps^T C \Gamma_{MN} \psi)^T = - \psi^T (-C \Gamma_{MN} C^{-1}) (-C) \eps = - \bar\psi\Gamma_{MN} \eps
\eea
where we used the 11d Majorana condition. We would first like to show a correspondence with the fermionic equation of motion and selfduality of $H_{MNP}$. In 6d, this correspondence is almost trivial to show. Namely, we have 
\bea
(\delta H_{MNP})^- &=& - \frac{i}{2} \nabla_Q \(\bar\psi \Gamma^Q \Gamma_{MNP} \eps\)
\eea
and by using the identity $\Gamma^Q \Gamma_{MNP} \Gamma_Q = 0$ and $\nabla_M \eps = \Gamma_M \eta$, we get
\bea
(\delta H_{MNP})^- &=& - \frac{i}{2} \nabla_Q \bar\psi \Gamma^Q \Gamma_{MNP} \eps
\eea
and we see that this variation vanishes on the fermionic equation of motion $\Gamma^M \nabla_M \psi = 0$. 

We would now like to show this correspondence between selfduality and the fermionic equation of motion again, but now in lightcone coordinates, following closely \cite{Lambert:2020scy}. To this end, we define
\bea
\G_{ij} &=& G_{ij} - r \sqrt{2} F_{i+} \kappa_j
\eea
where
\bea
G_{ij} &=& H_{ij+}\cr
F_{i+} &=& H_{i+-}
\eea
and we want to show that the selfdual part vanishes, $(\delta \G_{ij})^+ = 0$, on the fermionic equation of motion. So we first need to obtain the explicit expressions for the supersymmetry variation and for the fermionic equation of motion in lighcone coordinates. We begin with the supersymmetry variation. We have
\bea
\delta G_{ij} &=& - 2 i \nabla_i \(\bar\psi \Gamma_{j+} \eps\) - i \partial_+ \(\bar\psi \Gamma_{ij} \eps\)\cr
\delta F_{i+} &=& - i \partial_i \(\bar\psi\Gamma_{+-}\eps\) + i \partial_+ \(\bar\psi \Gamma_{i-} \eps\) - i \partial_- \(\bar\psi \Gamma_{i+} \eps\)
\eea
where $\nabla_i$ are 4d covariant derivatives. We expand 
\bea
\eps &=& e^{\frac{i \sqrt{2}}{r} x^-} \E + e^{-\frac{i \sqrt{2}}{r} x^-} \F\cr
\psi &=& e^{\frac{i \sqrt{2}}{r} x^-} \chi + e^{-\frac{i \sqrt{2}}{r} x^-} \zeta
\eea
where
\bea
\nabla_i \E &=& - \frac{3 i}{2} \kappa_i \E\cr
\partial_+ \E &=& - \frac{i}{r \sqrt{2}} \E
\eea
and corresponding relations for $\F$. We also expand 
\bea
\Gamma_{i\pm} &=& \t{\Gamma}_i \Gamma_{\pm} + \frac{r}{\sqrt{2}} \kappa_i \Gamma_{+-}\cr
\Gamma_{ij} &=& \t\Gamma_{ij} - r \sqrt{2} \kappa_i \t\Gamma_j \(\Gamma_+ - \Gamma_-\)
\eea
Then we get
\bea
\delta G_{ij} &=& - 2 i \nabla_i \(\bar\chi \t\Gamma_j \Gamma_+ \E\) - i \sqrt{2} r \nabla_i \(\bar\chi \Gamma_{+-} \E \kappa_j\)\cr
&& - i \partial_+ \(\bar\chi \t\Gamma_{ij} \E\) + i \sqrt{2} r \partial_+ \(\kappa_i \bar\chi \t\Gamma_j \Gamma_+ \E\)
\eea
We may now notice the appearance of a curly derivative from 
\bea
- 2 i \(\nabla_i - \frac{r}{\sqrt{2}} \kappa_i \partial_+\) \(\bar\chi \t\Gamma_j \Gamma_+ \E\) &=& - 2 i \D_i \(\bar\chi \t\Gamma_j \Gamma_+ \E\)
\eea
where we assume that $\partial_+ \kappa_i = 0$. So then we have
\bea
\delta G_{ij} &=& - 2 i \D_i \(\bar\chi \t\Gamma_j \Gamma_+ \E\) - i \sqrt{2} r \nabla_i \(\bar\chi \Gamma_{+-} \E \kappa_j\)\cr
&& - i \partial_+ \(\bar\chi \t\Gamma_{ij} \E\) 
\eea
We have 
\bea
\delta F_{i+} &=& - i \nabla_i \(\bar\chi \Gamma_{+-}\E\) + \frac{i r}{\sqrt{2}} \kappa_i \partial_+ \(\bar\chi \Gamma_{+-} \E\)
\eea
and then we get
\bea
\delta \G_{ij} &=& - 2 i \D_i \(\bar\chi \t\Gamma_j \Gamma_+ \E\) - i \sqrt{2} r \bar\chi \Gamma_{+-} \E \nabla_i \kappa_j\cr
&& - i \partial_+ \(\bar\chi \t\Gamma_{ij} \E\) 
\eea
or if we define
\bea
w_{ij} &=& \nabla_i \kappa_j - \nabla_j \kappa_i
\eea
then we can write this as
\bea
\delta \G_{ij} &=& - 2 i \D_i \(\bar\chi \t\Gamma_j \Gamma_+ \E\) - \frac{i r}{\sqrt{2}} \bar\chi \Gamma_{+-} \E w_{ij}\cr
&& - i \partial_+ \(\bar\chi \t\Gamma_{ij} \E\) 
\eea
We are now interested in extracting the selfdual part of this variation. To do this, we first recall the Weyl projection
\bea
\Gamma_- \E &=& 0
\eea
We have
\bea
\Gamma_{\pm} &=& \frac{1}{\sqrt{2}} \(\Gamma_t \pm \frac{1}{r} \Gamma_y\)\cr
\Gamma^{\pm} &=& \frac{1}{\sqrt{2}} \(\Gamma^t \pm r\Gamma^y\)
\eea
The Weyl projection can be written in the following alternative forms 
\bea
\Gamma^{\h{t}\h{y}} \E &=& \E\cr
\Gamma_{+-} \E &=& \E
\eea
Expressed in terms of 4d gamma matrices, we get
\bea
\delta \G_{ij} &=& 2 \sqrt{2} i \D_i \chi^* \gamma_j \E - \frac{i r}{\sqrt{2}} \chi^* \E w_{ij} - i \partial_+ \(\chi^* \gamma_{ij} \E\)
\eea
We can further write this as
\bea
\delta \G_{ij} &=& \frac{i}{\sqrt{2}} \D_k \chi^* [\gamma^k,\gamma_{ij}] \E - i \partial_+ \chi^* \gamma_{ij} \E - \frac{1}{r\sqrt{2}} \chi^* \gamma_{ij} \E - \frac{i r}{\sqrt{2}} \chi^* \E w_{ij}
\eea
Here we have rewritten this in terms of 6d Weyl components so that now all that remains of the $\Gamma_- \E = 0$ Weyl projection is 
\bea
\gamma \E &=& - \E
\eea
which amounts to that $\gamma_{ij} \E$ will be selfdual, and also $\gamma_{ij} \gamma_k \E$ will be antiselfdual simply because $\gamma_k \E$ is satisfying the opposite Weyl projection
\bea
\gamma \gamma_k \E &=& \gamma_k \E
\eea
as $\{\gamma_k,\gamma\} = 0$. Also since $w_{ij}$ is selfdual, we can now extract the selfdual part of the variation,
\bea
(\delta \G_{ij})^+ &=& \frac{i}{\sqrt{2}} \D_k \chi^* \gamma^k \gamma_{ij} \E - i \partial_+ \chi^* \gamma_{ij} \E - \frac{1}{r\sqrt{2}} \chi^* \gamma_{ij} \E - \frac{i r}{\sqrt{2}} \chi^* \E w_{ij}
\eea
We can also write this in the form
\bea
\delta \G_{ij} &=& - 2 \sqrt{2} i \F^* \gamma_j \D_i \zeta + \frac{i r}{\sqrt{2}} \F^* \zeta w_{ij} + i \partial_+ (\F^* \gamma_{ij} \zeta)
\eea
and then
\bea
(\delta \G_{ij})^+ &=& - \frac{i}{\sqrt{2}} \F^* \gamma_{ij} \gamma^k \D_k \zeta + \frac{i r}{\sqrt{2}} \F^* \zeta w_{ij} + i \F^* \gamma_{ij} \partial_+ \zeta + \frac{1}{r\sqrt{2}} \F^* \gamma_{ij} \zeta
\eea
that we can write as
\bea
&=& - \frac{i}{\sqrt{2}} \F^* \gamma_{ij} \(\gamma^k \D_k \zeta - \sqrt{2} \partial_+ \zeta + \frac{i}{r} \zeta\) + \frac{i r}{\sqrt{2}} \F^* \zeta w_{ij}
\eea
We now use the identity
\bea
\E &=& \frac{i r^2}{8} \gamma^{ij} \E w_{ij} 
\eea
to rewrite one term as
\bea
 - \frac{1}{r\sqrt{2}} \chi^* \gamma_{ij} \E &=& - \frac{i r}{8\sqrt{2}} \chi^* \gamma_{ij} \gamma^{kl} \E w_{kl}
\eea
and then we decompose
\bea
\gamma_{ij} \gamma^{kl} &=& \{\gamma_{ij},\gamma^{kl}\} - \gamma^{kl} \gamma_{ij}
\eea
Noting that $\{\gamma_{ij},\gamma^{kl}\} = - 8 \delta_{ij}^{kl}$ when acting on selfdual $w_{kl}$, the first term gives rise to a term
\bea
- \frac{i r}{\sqrt{2}} \chi^* \E w_{ij}
\eea
that cancels that corresponding term in $\delta \G_{ij}$, and we are left with 
\bea
\delta \G_{ij} &=& \frac{i}{\sqrt{2}} \D_k \chi^* [\gamma^k,\gamma_{ij}] \E - i \partial_+ \chi^* \gamma_{ij} \E + \frac{i r}{8\sqrt{2}} \chi^* \gamma^{kl} \gamma_{ij} \E 
\eea
and consequently
\bea
(\delta \G_{ij})^+ &=& \frac{i}{\sqrt{2}} \D_k \chi^* \gamma^k \gamma_{ij} \E - i \partial_+ \chi^* \gamma_{ij} \E + \frac{i r}{8\sqrt{2}} \chi^* \gamma^{kl} \gamma_{ij} \E 
\eea
We can write this as
\bea
(\delta \G_{ij})^+ &=& \frac{i}{\sqrt{2}} \(\D_k \chi^* \gamma^k - \sqrt{2} \partial_+ \chi^* + \frac{r}{8} \chi^* \gamma^{kl} w_{kl}\) \gamma_{ij} \E 
\eea
We now wish to show that this vanishes when the fermionic equation of motion is satisfied. Taking the complex conjugate of what is inside the parentesis, we get the requirement
\bea
\gamma^i \D_i \chi - \sqrt{2} \partial_+ \chi - \frac{r}{8} \gamma^{kl} \chi w_{kl} &=& 0
\eea
and indeed this is (a Weyl component of) the equation of motion.

Let us complete the supersymmetry variations. We have
\bea
\delta F_{i+} &=& - i \(\nabla_i \(\bar\chi \E\) - \frac{r}{\sqrt{2}} \kappa_i \partial_+ \(\bar\chi \E\)\)\cr
&=& - i \D_i \(\bar\chi \E\)\cr
&=& - i \D_i \(\chi^* \E\)
\eea
and, quite interestingly,
\bea
\delta \F_{ij} &=& - \frac{i r}{\sqrt{2}} \chi^* \E w_{ij}
\eea
This is interesting, because it is zero, up to a term that is proportional to $w_{ij}$. This is nothing like the usual supersymmetry variation, and in fact $\delta \D_i = 0$. And trivially
\bea
(\delta \F_{ij})^- &=& 0
\eea
since $w_{ij}$ is selfdual. We do not even need to use the fermionic equation of motion here.

We will now derive a 5d Lagrangian from the selfdual tensor field in 6d dimensions, following closely \cite{Lambert:2020scy}. We start by noting that 
\bea
H_{\h{i}\h{j}\h{\pm}} &=& E^i{}_{\h{i}} E^j{}_{\h{j}} \(H_{ij\pm} - r \sqrt{2} H_{i+-} \kappa_j\)\cr
H_{\h{i}\h{j}\h{k}} &=& E^i{}_{\h{i}} E^j_{\h{j}} E^k{}_{\h{k}} \(H_{ijk} - \frac{3r}{\sqrt{2}} H_{ij+} \kappa_k + \frac{3r}{\sqrt{2}} H_{ij-} \kappa_k\)\cr
H_{\h{i}\h{+}\h{-}} &=& E^i{}_{\h{i}} H_{i+-}
\eea
or if we define
\bea
F_{ij} &=& H_{ij-}\cr
G_{ij} &=& H_{ij+}\cr
F_{i+} &=& H_{i+-}
\eea
then 
\bea
H_{\h{i}\h{j}\h{+}} &=& E^i{}_{\h{i}} E^j{}_{\h{j}} \(G_{ij} - r \sqrt{2} F_{i+} \kappa_j\)\cr
H_{\h{i}\h{j}\h{-}} &=& E^i{}_{\h{i}} E^j{}_{\h{j}} \(F_{ij} - r \sqrt{2} F_{i+} \kappa_j\)\cr
H_{\h{i}\h{j}\h{k}} &=& E^i{}_{\h{i}} E^j_{\h{j}} E^k{}_{\h{k}} \(H_{ijk} + \frac{3r}{\sqrt{2}} \(F_{ij} - G_{ij}\) \kappa_k\)\cr
H_{\h{i}\h{+}\h{-}} &=& E^i{}_{\h{i}} F_{i+}
\eea
We have the Bianchi identity
\ben
3 \partial_{[i} H_{jk]+} - \partial_+ H_{ijk} &=& 0\label{Bianchiidentity}
\een
and we have the selfduality relation
\bea
H_{\h{i}\h{j}\h{k}} &=& \eps_{\h{i}\h{j}\h{k}}{}^{\h{l}\h{+}\h{-}} H_{\h{l}\h{+}\h{-}}
\eea
We define 
\bea
\eps_{\h{i}\h{j}\h{k}\h{l}\h{+}\h{-}} &=& \eps_{\h{i}\h{j}\h{k}\h{l}}
\eea
so we have
\bea
H_{\h{i}\h{j}\h{k}} &=& - \eps_{\h{i}\h{j}\h{k}}{}^{\h{l}} H_{\h{l}\h{+}\h{-}}
\eea
that we can write this as
\bea
H_{ijk} + \frac{3r}{\sqrt{2}} \(F_{ij} - G_{ij}\) \kappa_k + \eps_{ijk}{}^{l} F_{l+} &=& 0
\eea
The Bianchi identity (\ref{Bianchiidentity}) then becomes
\ben
3 \partial_{[i} G_{jk]} &=& - \partial_+ \(\frac{3r}{\sqrt{2}} \(F_{ij} - G_{ij}\) \kappa_k + \eps_{ijk}{}^l F_{l+}\)\label{B2}
\een
We define
\bea
\G_{ij} &=& G_{ij} - r \sqrt{2} F_{i+} \kappa_j\cr
\F_{ij} &=& F_{ij} - r \sqrt{2} F_{i+} \kappa_j
\eea
that enable us to express (\ref{B2}) in the following simple form
\bea
\eps^{ijkl} \D_{i} \G_{jk} &=& - 2 \partial_+ F^l{}_+
\eea
and from 
\bea
H_{\h{i}\h{j}\h{+}} &=& \frac{1}{2} \eps_{\h{i}\h{j}\h{+}}{}^{\h{k}\h{l}\h{+}} H_{\h{k}\h{l}\h{+}}
\eea
we get, by noting that $\eps_{\h{i}\h{j}\h{+}}{}^{\h{k}\h{l}\h{+}} = - \eps_{\h{i}\h{j}\h{+}\h{k}\h{l}\h{-}} =  - \eps_{\h{i}\h{j}\h{k}\h{l}\h{+}\h{-}} = - \eps_{\h{i}\h{j}\h{k}\h{l}}$, 
\bea
\G_{ij} &=& - \frac{1}{2} \eps_{ij}{}^{kl} \G_{kl}\cr
\F_{ij} &=& \frac{1}{2} \eps_{ij}{}^{kl} \F_{kl}
\eea
The next step will therefore be to replace straight capital letters with curly ones,
\bea
\partial_{[i} \(\G_{jk]} + r \sqrt{2} F_{j+} \kappa_k\) + \partial_+ \(\frac{r}{\sqrt{2}} \(\F_{ij} - \G_{ij}\) \kappa_k + \frac{1}{3} \eps_{ijk}{}^l F_{l+}\) &=& 0
\eea
because then we can dualize and get
\bea
- D_i \G^{il} + \frac{r}{\sqrt{2}} \eps^{ijkl} D_i \(F_{j+} \kappa_k\) + \frac{r}{\sqrt{2}} \kappa_i \partial_+ \(\F^{il} + \G^{il}\) + \partial_+ F^l{}_+ &=& 0
\eea
As a consequence of this equation, we have
\bea
- \(D_i \G^{il}\) \kappa_l + \partial_+ F^l{}_+ \kappa_l &=& 0
\eea
that we can also write as
\bea
- D_i \(\G^{ij} \kappa_j\) + \frac{1}{2} G^{il} w_{il} + \partial_+ F^l{}_+ \kappa_l &=& 0
\eea
but the second term is vanishing, as one can see by replacing $G^{il}$ with $\G^{il}$ whcih is antiselfdual so contracting with a selfdual $w_{il}$ gives zero. And moreover $\kappa^l w_{il} = 0$. So we have 
\bea
D_i  \(\G^{ij} \kappa_j\) &=& \kappa_i \partial_+ F^i{}_+  
\eea
which will be a useful relation that we will use later. We may also write 
\bea
- \D_i \G^{il} + \frac{r}{\sqrt{2}} \eps^{ijkl} D_i \(F_{j+} \kappa_k\) + \frac{r}{\sqrt{2}} \kappa_i \partial_+ \F^{il} + \partial_+ F^l{}_+ &=& 0
\eea

We have the Bianchi identity
\bea
3 \partial_{[i} H_{jk]-} &=& \partial_- H_{ijk}
\eea
but if we put $\partial_- = 0$ upon dimensional reduction, then this reduces to 
\bea
\eps^{ijkl} \partial_{i} F_{jk} &=& 0
\eea
Again replacing straight capital $F$ with curly $\F$, we first get
\bea
\eps^{ijkl} \partial_{i} \(\F_{jk} + r \sqrt{2} F_{j+} \kappa_k\) &=& 0
\eea
and then by using selfduality this becomes
\bea
D_i \F^{il} + \frac{r}{\sqrt{2}} \eps^{ijkl} D_i \(F_{j+} \kappa_k\) &=& 0
\eea
But the nicest way to express this same equation is as
\bea
\eps^{ijkl} \D_{i} \F_{jk} &=& 0
\eea

We have the Bianchi identity
\bea
2\partial_{[i} F_{j]+} + \partial_+ F_{ij} &=& 0
\eea
Replacing $F$ with $\F$ it becomes
\bea
2\D_{[i} F_{j]+} + \partial_+ \F_{ij} &=& 0
\eea

Finally, we return to 
\bea
H_{ijk} + \frac{3r}{\sqrt{2}} \(F_{ij} - G_{ij}\) \kappa_k + \eps_{ijk}{}^{l} F_{l+} &=& 0
\eea
and apply the Bianchi identity $\eps^{ijkl} \partial_l H_{ijk} = 0$. We then get
\bea
\D^i F_{i+} &=& \frac{r}{2\sqrt{2}} \F_{ij} w^{ij}
\eea
where we define 
\bea
\D_i &=& D_i - \frac{r}{\sqrt{2}} \kappa_i \partial_+
\eea

We have thus got two types of equations of motion,
\bea
- \D_i \G^{il} + \frac{r}{\sqrt{2}} \eps^{ijkl} D_i \(F_{j+} \kappa_k\) + \frac{r}{\sqrt{2}} \kappa_i \partial_+ \F^{il} + \frac{1}{2} \partial_+ F^l{}_+ &=& 0\cr
\D^i F_{i+} - \frac{r}{2\sqrt{2}} \F_{ij} w^{ij} &=& 0
\eea
and in addition to these, we have the selfduality equations
\bea
\G_{ij} &=& - \frac{1}{2} \eps_{ij}{}^{kl} \G_{kl}\cr
\F_{ij} &=& \frac{1}{2} \eps_{ij}{}^{kl} \F_{kl}
\eea
The equation 
\ben
D_i \F^{il} + \frac{r}{\sqrt{2}} \eps^{ijkl} D_i \(F_{j+} \kappa_k\) &=& 0\label{extra}
\een
surely looks very much like an independent equation of motion, but actually it is not. It is a direct consequence of $\eps^{ijkl} \partial_i F_{jk} = 0$ together with the selfduality equation of motion for $\F_{ij}$. That means we do not need to demand that the equation
(\ref{extra}) follows from an action upon the variation of a gauge field as one normally would expect. Now one may ask some questions about number of components. Let us be very brief and just notice that selfdual $H_{MNP}$ has $10$ components, just as do selfdual $F_{ij}$ and $F_{i+}$ together, as $6 + 4 = 10$. So we do not expect $G_{ij}$ shall be part of the supermultiplet upon dimensional reduction. Only $F_{ij}, F_{i+}$ should be part of the vector muliplet. It then seems reasonable to assume that the antiselfdual $\G_{ij}$ shall be viewed as a Lagrange multiplier field that is imposing selfduality on $\F_{ij}$, rather than as a dynamical field that contributes to additional degrees of freedom. We now make the following ansatz for a gauge field Lagrangian,
\bea
\L_A &=& b \F^{ij} \G_{ij} + c F^i{}_+ F_{i+} + d \eps^{ijkl} F_{ij} F_{k+} \kappa_l + e \eps^{ijkl} G_{ij} F_{k+} \kappa_l
\eea
and treat $\G_{ij}$ (assumed to be antiselfdual from the outset), $A_i$ and $A_+$ as independent fields that we shall vary to derive the classical equations of motion. Then these equations of motion become
\bea
\F_{ij} - \frac{1}{2} \eps_{ij}{}^{kl} \F_{kl} &=& 0\cr
\(b r \sqrt{2} + 2 e\) D_i \(\G^{ij} \kappa_j\) - 2 c D^i F_{i+} - d \F_{ij} w^{ij} &=& 0\cr
- 2 b D_i \G^{im} + \(b r \sqrt{2} + 2 e\) \kappa_i \partial_+ \G^{i m} + 2 d \eps^{mikl} D_i \(F_{k+} \kappa_l\) + 2 d \partial_+ \F^{mj} \kappa_j + 2 c \partial_+ F^m{}_+  &=& 0
\eea
We now write the second relation as
\bea
- 2 c D_i F^i{}_+ + \( b r \sqrt{2} + 2 e\) \kappa_i \partial_+ F^i{}_+ - d \F_{ij} w^{ij} &=& 0
\eea
By now requiring the combination $\D_i = D_i - \frac{r}{\sqrt{2}} \kappa_i \partial_+$ to appear, we get the following equations
\bea
\frac{br \sqrt{2} + 2 e}{2c} &=& \frac{r}{\sqrt{2}}\cr
\frac{br \sqrt{2} + 2 e}{2b} &=& \frac{r}{\sqrt{2}}\cr
\frac{d}{b} &=& - \frac{r}{\sqrt{2}}\cr
\frac{c}{b} &=& 1
\eea
These equations have the following unique solution 
\bea
b &=& 1\cr
c &=& 1\cr
d &=& - \frac{r}{\sqrt{2}}\cr
e &=& 0
\eea
up to one overall constant. Fixing that overall constant to be $1/4$, the Lagrangian is given by
\bea
\L_A &=& \frac{1}{4} \(\F^{ij} \G_{ij} + F^i{}_+ F_{i+} - \frac{r}{\sqrt{2}} \eps^{ijkl} \F_{ij} F_{k+} \kappa_l\)
\eea
where we have also replaced $F$ with $\F$ in the graviphoton term, which we can do freely by just noting that $\kappa_j \kappa_l = 0$ upon antisymmetrization in $j$ and $l$. The supersymmetry variation of this Lagrangian is  
\bea
\delta \L_A &=& - \frac{i}{\sqrt{2}} \chi^* \gamma_j \E \D_i \F^{ij} + \frac{i}{4} \chi^* \gamma^{ij} \E \partial_+ \F_{ij} + \frac{i}{2} \chi^* \E \D^i F_{i+}\cr
&& - \frac{ir}{4\sqrt{2}} \chi^* \E \G^{ij} w_{ij} - \frac{ir}{2\sqrt{2}} \chi^* \E \F^{ij} w_{ij}
\eea
The fourth term is identically zero because $\G_{ij}$ is antiselfdual off-shell. 

Next, we obtain the supersymmetry variation
\bea
\delta \psi &=& \frac{1}{12} \Gamma^{MNP} \eps H_{MNP}
\eea
in 4d. To this end, it is advantageous to first recast this in flat space indices,
\bea
\delta \psi &=& \frac{1}{4} \Gamma^{\h{i}\h{j}\h{+}} \eps H_{\h{i}\h{j}\h{+}} + \frac{1}{4} \Gamma^{\h{i}\h{j}\h{-}} \eps H_{\h{i}\h{j}\h{-}} + \frac{1}{2} \Gamma^{\h{i}\h{+}\h{-}} \eps H_{\h{i}\h{+}\h{-}} 
\eea
Then it immediately follows that
\bea
\delta \psi &=& \frac{1}{4} \t\Gamma^{ij} \Gamma^{\h{-}} \eps \F_{ij} + \frac{1}{2} \t\Gamma^i \Gamma^{\h{+}\h{-}} \eps F_{i+}
\eea
which in terms of 4d gamma matrices reads
\bea
\delta \chi &=& \frac{1}{2\sqrt{2}} \gamma^{ij} \E \F_{ij} - \frac{1}{2} \gamma^i \E F_{i+}
\eea
Then let us look at each term in turn in the fermionic action
\bea
\L_F &=& \frac{i}{2} \chi^* \gamma^i \D_i \chi - \frac{i}{\sqrt{2}} \chi^* P_- \partial_+ \chi\cr
&& + \frac{1}{r} \chi^* P_+ \chi - \frac{i r}{16} \chi^* \gamma^{ij} P_- \chi w_{ij}
\eea
The variation of the first two derivative terms becomes after using two types of Bianchi identities 
\bea
\delta \(\L_F^I + \L_F^{II}\) &=& \frac{i}{\sqrt{2}} \chi^* \gamma^i \E \D^j \F_{ji} - \frac{1}{2} \chi^* \E \D^i F_{i+} - \frac{i}{4} \chi^* \gamma^{ij} \E \partial_+ \F_{ij}\cr
&& - \frac{1}{r 2 \sqrt{2}} \chi^* \gamma^{ij} \E \F_{ij}
\eea
The first line is exactly canceling corresponding terms in $\delta \L_A$. The variation of the two last mass terms gives
\bea
\delta \(\L_F^{II} + \L_F^{IV}\) &=& - \frac{1}{r} \chi^* \gamma^i \E F_{i+} - \frac{i r}{16 \sqrt{2}} \chi^* \gamma^{ij} \gamma^{kl} \E w_{ij} \F_{kl}
\eea
Ideally we had wanted these to cancel against the last term in $\delta\L_A$,
\bea
\delta \L_A^{VI} &=& - \frac{i r}{2\sqrt{2}} \chi^* \E w_{ij} \F^{ij}
\eea
We do not seem to get a perfect cancelation, but let us note that we can rewrite the last term in $\delta \(\L_F^{II} + \L_F^{IV}\)$ as
\bea
- \frac{i r}{16 \sqrt{2}} \chi^* \(\{\gamma^{ij},\gamma^{kl}\} - \gamma^{kl} \gamma^{ij}\) \E w_{ij} \F_{kl} &=& \frac{i r}{2 \sqrt{2}} \chi^* \E w_{ij} \F^{ij} + \frac{i r}{16 \sqrt{2}} \chi^* \gamma^{kl} \gamma^{ij} \E w_{ij} \F_{kl} \cr
&=& \frac{i r}{2 \sqrt{2}} \chi^* \E w_{ij} \F^{ij} + \frac{1}{r 2 \sqrt{2}} \chi^* \gamma^{ij} \E \F_{ij}
\eea
The first term cancels against $\delta \L_A^{VI}$ and the second term cancels the last term in $\delta \(\L_F^I + \L_F^{II}\)$. The final result is that we have the following nonzero variation of the Lagrangian,
\ben
\delta \L &=& - \frac{1}{r} \chi^* \gamma^i \E F_{i+}\label{unwanted}
\een
Since the 6d metric inverse $g^{ij}$ is equal to the 4d metric inverse $G^{ij}$ and since the index $i$ in $F_{i+} = H_{i+-}$ can be extended to indices $+$ and $-$ without changing anything since $H_{++-}$ and $H_{-+-}$ are zero anyway, we can view $i$ as a 6d index contracted by the 6d metric. This means that we can write this result in terms of 6d flat space indices as
\bea
\delta \L &=& \frac{i}{r\sqrt{2}} \delta B^{\h{i}}{}_{\h{+}} H_{\h{i}\h{+}\h{-}}
\eea
and by using the selfduality relation
\bea
H_{\h{i}\h{j}\h{k}} &=& - \eps_{\h{i}\h{j}\h{k}}{}^{\h{l}} F_{\h{l}\h{+}\h{-}}
\eea
we can further write this as
\bea
\delta \L &=& \frac{i}{r\sqrt{2}} \eps^{\h{i}\h{j}\h{k}\h{l}\h{+}\h{-}} \delta B_{\h{i}\h{+}} H_{\h{j}\h{k}\h{l}} 
\eea
Now we can change to 6d curved space indices and then this becomes
\bea
\delta \L &=& - \frac{i}{r\sqrt{2}} \eps^{ijkl} H_{ijk} \delta B_{l+} 
\eea
where we define $\eps^{ijkl} = - \eps^{ijkl+-}$. We now wish to show that this can be expressed as a total variation of some topological term of the form
\bea
\L_{top} &=& \eps^{ijkl} H_{ijk} B_{l+}
\eea
up to some constant factor. When we expand its variation, we find two types of terms,
\bea
\delta \L_{top} &=& 3 \eps^{ijkl} \partial_i \delta B_{jk} B_{l+} +\eps^{ijkl} H_{ijk} \delta B_{l+}
\eea
The first term here can be further written as
\bea
- 3 \eps^{ijkl} \delta B_{jk} \partial_i B_{l+} &=& - \frac{3}{2} \eps^{ijkl} \delta B_{jk} H_{il+}
\eea
where we dropped a couple of total derivative terms. Now, if we change to flat space indices we see the emergence of an antiselfdual $H_{\h{i}\h{j}\h{+}} = \G_{\h{i}\h{j}}$ and so what this term becomes is something that is proportional to $\delta B^{ij} \G_{ij}$ and this is zero, because $\delta B^{ij} \sim \chi^* \gamma^{ij} \E$ and we have that $\gamma^{ij} \E \G_{ij} = 0$ since $\G_{ij}$ is antiselfdual and $\E$ is Weyl. One way to see this is by noting that $\gamma^{ij} \E w_{ij}$ is nonzero where $w_{ij}$ is selfdual. This means that we are left with only the second term,
\bea
\delta \L_{top} &=& \eps^{ijkl} H_{ijk} \delta B_{l+}
\eea
as we wanted to show. So by adding the topological term
\bea
\L_{top} &=& \frac{i}{r\sqrt{2}} \eps^{ijkl} H_{ijk} B_{l+}
\eea
we find that its variation cancels the variation $\delta \L$ in (\ref{unwanted}) above.

Let us now study the matter part supersymmetry. The Lagrangian is
\bea
\L_F &=& \frac{i}{2} \chi^{\dag} \gamma^i D_i \chi - \frac{i}{\sqrt{2}} \chi^{\dag} P_- \partial_+ \chi + \frac{1}{r} \chi^{\dag} P_+ \chi\cr
&& - \frac{i}{2r} \chi \gamma^{12} P_- \chi - \frac{i r}{2\sqrt{2}} \kappa_i \chi^{\dag} \gamma^i \partial_+ \chi\cr
&& - \frac{1}{2} \kappa_i \chi^{\dag} \gamma^i \chi
\eea
The supersymmetry variation is 
\bea
\delta \chi &=& - \gamma^i \tau^A \E \D_i \phi^A - \frac{2 i}{r} \tau^A \E \phi^A
\eea
where we define 
\bea
\D_i &=& D_i - \kappa_i \partial_y\cr
\partial_y &=& \frac{r}{\sqrt{2}} \(\partial_+ - \partial_-\)
\eea
Using this generalized derivative on the fermion, and the expansion where 
\bea
\partial_- \chi &\rightarrow & \frac{i \sqrt{2}}{r} \chi
\eea
we find that the Lagrangian simplifies to 
\bea
\L_F &=& \frac{i}{2} \chi^{\dag} \gamma^i \D_i \chi - \frac{i}{\sqrt{2}} \chi^{\dag} P_- \partial_+ \chi \cr
&& + \frac{1}{r} \chi^{\dag} P_+ \chi - \frac{i}{2r} \chi \gamma^{12} P_- \chi 
\eea
We thus need to carefully define the operator $\D_i$ acting on bosons and fermions respectively, as
\bea
\D_i \phi &=& \partial_i \phi - \frac{r}{\sqrt{2}} \partial_+ \phi\cr
\D_i \chi &=& D_i \chi - \frac{r}{\sqrt{2}} \kappa_i \partial_+ \chi + i \kappa_i \chi
\eea
Similarly then when this generalized derivative acts on the supersymmetry parameter, and then one finds the following Killing spinor equation
\bea
\D_i \E &=& 0
\eea
We get the supersymmetry variation
\bea
\delta \L_F &=& - \frac{i}{2} \chi^{\dag}\gamma^{ij}\tau^A\E[\D_i,\D_j]\phi^A - i \chi^{\dag}\tau^A\E\D_i^2\phi^A\cr
&& - \frac{2\sqrt{2}}{r} \chi^{\dag}\tau^A\E\partial_+\phi^A\cr
&& + \frac{2i}{r^2} \chi^{\dag}\tau^A\E\phi^A - \frac{2}{r} \chi^{\dag}\gamma^{12}\tau^A\E\phi^A
\eea
Two terms cancel by using 
\bea
[\D_i,\D_j] \phi &=& - \frac{r}{\sqrt{2}} w_{ij} \partial_+ \phi
\eea
and 
\bea
\gamma^{12} \E &=& - i \E\cr
\gamma^{34} \E &=& - i \E
\eea
and we get
\bea
\delta \L_F &=& - i \chi^{\dag}\tau^A\E\D_i^2\phi^A\cr
&& + \frac{4i}{r^2} \chi^{\dag}\tau^A\E\phi^A 
\eea
Let us now turn to the scalar fields' Lagrangian
\bea
\L_S &=& - \frac{1}{2} (\D_i \phi^A)^2 - \frac{2}{r^2} (\phi^A)^2
\eea
Using the variation
\bea
\delta \phi^A &=& i \chi^{\dag}\tau^A\E
\eea
we find that $\delta \L_S + \delta \L_F = 0$.

Before turning to the non-Abelian case, let us first summarize the Abelian case. We have the Lagrangian
\bea
\L &=& \L_A + \L_{matter} + \L_{top}
\eea
where 
\bea
\L_A &=& \frac{1}{4} \(\F^{ij} \G_{ij} + F^i{}_+ F_{i+} - \frac{r}{\sqrt{2}} \eps^{ijkl} \F_{ij} F_{k+} \kappa_l\)\cr
\L_{matter} &=& \frac{i}{2} \chi^* \gamma^i \D_i \chi - \frac{i}{\sqrt{2}} \chi^* P_- \partial_+ \chi\cr
&& + \frac{1}{r} \chi^* P_+ \chi - \frac{i r}{16} \chi^* \gamma^{ij} P_- \chi w_{ij}\cr
&& - \frac{1}{2} (\D_i \phi^A)^2 - \frac{2}{r^2} (\phi^A)^2\cr
\L_{top} &=& \frac{i}{r \sqrt{2}} \eps^{ijkl} H_{ijk} B_{l+}  
\eea
and the supersymmetry variations
\bea
\delta \phi^A &=& i \chi^* \tau^A\E\cr
\delta \chi &=& \frac{1}{2\sqrt{2}} \gamma^{ij} \E \F_{ij} - \frac{1}{2} \gamma^i \E F_{i+} - \gamma^i \tau^A \E \D_i \phi^A - \frac{2 i}{r} \tau^A \E \phi^A\cr
\delta A_i &=& - \frac{i r}{\sqrt{2}} \kappa_i \chi^* \E\cr
\delta A_+ &=& - i \chi^* \E\cr
\delta F_{i+} &=& - i \D_i \(\chi^* \E\)\cr
\delta \F_{ij} &=& - \frac{i r}{\sqrt{2}} \chi^* \E w_{ij}\cr
\delta \G_{ij} &=& 2 \sqrt{2} i \D_i \chi^* \gamma_j \E - \frac{i r}{\sqrt{2}} \chi^* \E w_{ij} - i \partial_+ \(\chi^* \gamma_{ij} \E\)
\eea
To see whether a non-Abelian generalization is possible, let us start by replacing all derivatives with gauge covariant derivatives,
\bea
\D_i \phi^A &=& D_i \phi^A - \kappa_i D_y \phi^A\cr
D_i \phi^A &=& \partial_i \phi^A - i e [A_i,\phi^A]\cr
D_y \phi^A &=& \partial_y \phi^A - i e [A_y,\phi^A]
\eea
in the supersymmetry variations. Then by noting that 
\bea
[\D_i,\D_j] \phi^A &=& - i e [\F_{ij},\phi^A]
\eea
we get
\bea
\delta \L &=& - \frac{e}{2} \chi^* \gamma^{ij} \tau^A \E [\F_{ij},\phi^A]
\eea
To cancel this variation, one might be tempted to add the following term to the Lagrangian,
\bea
\Delta \L &=& \frac{e}{\sqrt{2}} \chi^* \tau^A [\chi,\phi^A]
\eea
But if we do that, then that term will upon a supersymmetry variation generate a host of new terms, such as
\ben
\chi^* \gamma^i \E [\D_i \phi^A,\phi^A]\label{u}
\een
but we can not cancel this term by anything. The only candidate term $(\D_i \phi^A)^2$ does not work because the supersymmetry variation of the gauge potential $\A_i$ is vanishing, so it can not give rise to something that is proportional to $\chi^* \gamma_i \E$. So we can not cancel the variation (\ref{u}) and therefore we shall not add any extra commutator terms to the Lagrangian. 

Instead we shall modify the supersymmetry variation of $\G_{ij}$ by adding a term\footnote{This is in accordance with the Lambert-Papageorgakis theory, where 
\bea
\delta H_{MNP} &\sim & .. + [\phi^A,\bar\psi] \Gamma^A \Gamma_{MNPQ} \eps v^Q 
\eea
if we notice that the only surviving combination of gamma matrices can be $\Gamma_{ij+-}$, which simply means that the commutator only enters in $H_{ij+}$, or in other words $\G_{ij}$.}
\bea
\Delta \delta \G_{ij} &=& \frac{e}{2} [\chi^* \gamma_{ij} \tau^A \E,\phi^A]
\eea

If we could make the gauge choice $A_+ = 0$ and then just forget about $\delta A_+$ altogether, then since $\delta \A_i = 0$, we would have no cubic term in the fermionic fields that could appear when we vary the gauge potential in the fermionic kinetic term. But imposing the gauge choice $A_+ =0$ is unsatisfactory since this gauge choice breaks supersymmetry by itself. We can avoid this problem of gauge fixing by reducing supersymmetry by another half. We then impose the Weyl projection 
\bea
\tau^5 \E &=& \E
\eea
Then we have the supersymmetry variation
\bea
\delta \phi^5 &=& i \chi^* \E
\eea
and we see that the combination $A_+ - \phi^5$ is a supersymmetric invariant,
\bea
\delta \(A_+ - \phi^5\) &=& 0
\eea
We then obtain a supersymmetric Lagrangian by simply adding commutator terms that involve $\phi^5$ for each place where there is a gauge field $A_+$. Such commutator terms are of course gauge invariant by themselves. But we can repackage these terms into a new derivative 
\bea
\D_+ &=& D_+ + i e [\phi^5,\bullet]
\eea
where $D_+ = \partial_+ - i e [A_+,\bullet]$. One may worry that ordinary derivative acts on a fermionic field, but that is just because of how we have set up our Lagrangian. We have already taken into account all those curvature corrections when we analysed the Abelian case and those curvature corrections will not be affected in any significant way by the non-Abelian generalization. We now obtain a full supersymmetric non-Abelian Lagrangian by replacing every occurence of $\partial_+$ with $\D_+$ as we defined it above (with an ordinary derivative $\partial_+$ rather than a curvature covariant $\nabla_+$). There is now at this stage no need to  impose any gauge fixing condition on $A_+$. 

The $B\wedge H$ term is straightforwardly generalized to the non-Abelian case as $B^a \wedge H^a$ where $a$ is the adjoint gauge group index. The supersymmetry variation of $\t{B}_{i+}$ is similarly generalized by just attaching that adjoint gauge group index as $\delta{B}^a_{i+} = - i \sqrt{2} (\chi^a)^* \gamma_i \E$. We also assume the duality relation is generalized to the non-Abelian case as $H_{ijk}^a = \eps_{ijkl} F_{l+}^a$. 

As we did not put any component of the fermionic field to zero here, as we did for the case of time reduction, we do not expect our 5d Lagrangian will be possible to derive by turing on an R-gauge field in some dual formulation. In particular, we do not expect the closure relations will be of a standard form.

\section*{Acknowledgements}
I would like to thank Dongsu Bak for discussions. This work was supported in part by NRF Grant 2020R1A2B5B01001473 and NRF Grant 2020R1I1A1A01052462.

\appendix
\section{A 6d formulation of 5d SYM}\label{A}
There is a 6d formulation of 5d SYM where one introduces a vector field $v^M$ and requires all fields to have vanishing Lie derivatives along that vector field \cite{Lambert:2010wm}, \cite{Gustavsson:2018rcc}, \cite{Gustavsson:2020ugb}. We did not make explicit use of this 6d formulation of 5d SYM. But it was this formulation that originally motivated us to search for consistent supersymmetric truncations, and the two cases that we have studied in this paper can be at least intuitvely quite clearly understood by looking at this formulation of the theory where they emerge as the Weyl projections (\ref{W1}) and (\ref{W2}) respectively. 

The 6d supersymmetry variations look like a non-Abelian generalization of the Abelian M5 brane, but of course there is a catch. Namely we do not have closure relations satisfied for these variations, unless two terms vanish, namely the terms in (\ref{first}) and (\ref{second}). Let us present this in detail. The supersymmetry variations are given by
\bea
\delta \phi^A &=& i \bar\eps\Gamma^A\psi\cr
\delta H_{MNP} &=& 3 i D_P\(\bar\eps\Gamma_{MN}\psi\)+e\bar\eps\Gamma_{MNPQ}\Gamma^A[\psi,\phi^A]v^Q\cr
\delta A_N &=& i \bar\eps\Gamma_{NP}\psi v^P\cr
\delta \psi &=& \frac{1}{12}\Gamma^{MNP}\eps H_{MNP} + \Gamma^M\Gamma^A\eps D_M\phi^A - 4\Gamma^A\eta\phi^A - \frac{ie}{2}\Gamma_M\Gamma^{AB}\eps[\phi^A,\phi^B]v^M
\eea
Here 
\bea
D_M \phi^A &=& \partial_M\phi^A - i e [A_M,\phi^A] + V_M^{AB}\phi^B
\eea
where $V_M$ is an R-gauge field, and 
\ben
D_M \eps &=& \Gamma_M \eta - \frac{1}{8} \Gamma^A \Gamma^{RST} \Gamma_M\eps T^A_{RST}\cr
D_M \bar\eps &=& - \bar\eta \Gamma_M - \frac{1}{8} \bar\eps \Gamma_M \Gamma^{RST} \Gamma^A T^A_{RST}\label{CKE}
\een
Here $v^M$ is a Killing vector field and $\L_v$ denotes the Lie derivative along this Killing vector field. We will impose the gauge condition 
\ben
A_M v^M &=& 0\label{Av}
\een
which is a very natural gauge condition if we think on $A_M$ as $B_{MN} v^N$. Now this correspondence is at present unknown to us for the nonabelian case where $H_{MNP}$ is all that we have. We would like to know how to express the theory in terms of some nonabelian gauge potential $B_{MN}$ but at present we do not have such a formulation. Nevertheless, the gauge potential will be assumed to satisfy the gauge condition (\ref{Av}).

We define the 6d chirality matrix 
\bea
\Gamma &=& \Gamma^{\h{0}\h{1}\h{2}\h{3}\h{4}\h{5}}
\eea
in flat tangent space and we assume that spinor and supersymmetry parameter have opposite chiralities
\bea
\Gamma \psi &=& \psi\cr
\Gamma \eps &=& -\eps
\eea
We use 11d gamma matrices where $\Gamma_M$ denote spacetime gamma matrices for $M= 0,1,...,5$ and $\Gamma_A$ denote five transverse space gamma matrices for $A=1,2,3,4,5$ and these anticommute, $\{\Gamma_M,\Gamma_A\} = 0$. 

For the closure computation of these supersymmetry variations, we define 
\bea
S^M &=& \bar\eps\Gamma^M \eps
\eea
and the gauge parameter 
\bea
\Lambda &=& - i \bar\eps\Gamma_Q\Gamma^A\eps\phi^A v^Q
\eea
and we assume that $\eps$ is a commuting spinor, since that simplifies the closure computation a bit yet without imposing any restrictions.

The superconformal algebra in curved space is 
\bea
\delta_{\eps}^2 &=& - i \L_S - 2 i \W - 2 \bar\eps\Gamma^{AB}\eta R^{AB} + \delta_{gauge}
\eea
where
\bea
R^{AB} &=& \frac{i}{2} \Gamma^{AB}\cr
R^{AB}_{CD} &=& 2 i \delta^{AB}_{CD}
\eea
and 
\bea
\W_{\phi^A} &=& 2\cr
\W_{\psi} &=& \frac{5}{2}\cr
\W_{H_{MNP}} &=& 0
\eea
are the Weyl weights.

As always with closure relations, we can express these in terms of conventional Lie derivatives $\L_S$, or in terms of gauge covariant Lie derivatives $\mb{L}_S$. These are related as
\bea
- i \mb{L}_S \phi^A &=& - i \L_S \phi^A - i e [\phi^A,\Delta \lambda] - i S^M V_M^{AB} \phi^B\cr
- i \mb{L}_S \psi &=& - i \L_S \psi - i e [\psi,\Delta \lambda] - \frac{i}{4} S^M V_M^{AB} \Gamma^{AB} \psi\cr
- i \mb{L}_S H_{MNP} &=& - i \L_S H_{MNP} - i e [H_{MNP},\Delta \lambda]\cr
- i \mb{L}_S A_M &=& - i S^N F_{NM} \cr
&=& - i \L_S A_M + D_M \(\Delta \lambda\)
\eea
where $\Delta \lambda = i S^M A_M$. We thus see that from $\mb{L}_S$ we get $\L_S$ plus some extra gauge transformation and R-rotation. 

If we assume that $\eps$ is commuting, then we find the following closure relations,
\bea
\delta^2 \phi^A &=& - i \mb{L}_S \phi^A - 4 i \bar\eps\eta \phi^A - 4 i \bar\eps\Gamma^{AB}\eta \phi^B - i e [\phi^A,\Lambda]\cr
\delta^2 H_{MNP} &=& - i \mb{L}_S H_{MNP} \cr
&& + 3 i D_M\(S^T \(H_{NPT}^- - 6\phi^A T^A_{NPT}\)\)\cr
&& - 4 i S^T D_{[P} H_{MNT]} - e \eps_{MNPQRS} \bar\eps  \Gamma^S\eps [D_R \phi^A,\phi^A] v^Q - \frac{ie}{2}\bar\eps\Gamma^V\eps \eps_{MNPQUV} \bar\psi^b \Gamma^U\psi^a v^Q\cr
&& + 3 e \bar\eps \Gamma_P \Gamma^A \eps [H_{MNQ}v^Q-F_{MN},\phi^A]\cr
&& - i e [H_{MNP},\Lambda]\cr
&& + \frac{e}{2} \L_v\(\bar\eps\Gamma^{AB}\Gamma_{MNP}\eps\)[\phi^A,\phi^B]\cr
\delta^2 \psi &=& - i \mb{L}_S \psi - 2 i \bar\eta\eps \frac{5}{2} \psi - i \bar\eta\Gamma^{AB}\eps \Gamma^{AB} \psi - i e [\psi,\Lambda]\cr
&& + \frac{3i}{8} S^Q \Gamma_Q \(\Gamma^P D_P \psi + \frac{1}{4}\Gamma^{RST}\Gamma^A\psi T^A_{RST} - i e \Gamma_M \Gamma^A [\psi,\phi^A] v^M\)\cr
&& + 2 i c^{QB} \Gamma_Q \Gamma^B \(\Gamma^P D_P \psi + \frac{1}{4}\Gamma^{RST}\Gamma^A\psi T^A_{RST} - i e \Gamma_M \Gamma^A [\psi,\phi^A] v^M\)\cr
\delta^2 A_M &=& -i \mb{L}_S A_M + D_M\(-i\bar\eps\Gamma_N\Gamma^A\eps\phi^A v^N\) \cr
&& + i S^T F_{TM} - i S^T \(H_{MNT}^+ + 6 T^A_{MNP}\phi^A\) v^N\cr
&& + i \L_v\(\bar\eps\Gamma_M\Gamma^A \eps \phi^A\) 
\eea
Apart from the term
\ben
\frac{e}{2} \L_v\(\bar\eps\Gamma^{AB}\Gamma_{MNP}\eps\)[\phi^A,\phi^B]\label{first}
\een
in $\delta^2 H_{MNP}$ and the term
\ben
i \L_v\(\bar\eps\Gamma_M\Gamma^A \eps \phi^A\) \label{second}
\een
in $\delta^2 A_M$, we can now obtain closure up to a gauge transformation with gauge parameter $\lambda = - i \bar\eps\Gamma_M\Gamma^A\eps \phi^A v^M$ if certain equations of motion are satisfied. Closure on $H_{MNP}$ requires the following equations of motion,
\ben
H_{NPT}^- - 6\phi^A T^A_{NPT} &=& 0\label{a}\\
H_{MNQ}v^Q-F_{MN} &=& 0\label{b1}
\een
and closure on $A_M$ requires the equation of motion
\ben
F_{TM} - \(H_{MNT}^+ + 6 T^A_{MNP}\) v^N &=& 0\label{c}
\een
By adding $0 = H_{NPT}^- - 6\phi^A T^A_{NPT}$ to $H_{MNT}^+ + 6 T^A_{MNP} \phi^A$ we get $H_{MNT} = H_{MNT}^+ + H_{MNT}^-$ and (\ref{c}) reduces to (\ref{b1}). Of course the presence of the terms (\ref{first}) and (\ref{second}) means that these 6d supersymmetry variations do not close, unless both these terms vanish. One way to make these two terms vanish is by requiring the Lie derivative vanishes on every field and also on the supersymmetry parameter, $\L_v \eps = 0$, where $\L_v$ denotes the Lie derivative along $v^M$. This is the usual dimensional reduction along the vector field $v^M$. 

Could there be some other ways to achieve closure? At least for the first term (\ref{first}), we can make that term disappear without requiring $\L_v \eps = 0$. To see this more clearly, let us notice that a corresponding commutator term sits in the supersymmetry variation of the $(2,0)$ tensor multiplet fermion $\psi$ as
\bea
\delta \psi &=& ... - \frac{i e}{2} \Gamma_M \Gamma^{AB} \eps [\phi^A,\phi^B] v^M
\eea
and here we can see two ways for this commutator term to vanish. 

One is by just keeping one scalar field, say $\phi^5$ and reduce supersymmetry by imposing the R-symmetry Weyl projection 
\ben
\Gamma^{A=5} \eps &=& \eps\label{W1}
\een
and discarding the hypermultiplet. Of course, with just one scalar field, there will be no nontrivial commutator term $[\phi^A,\phi^B]$, but having to discard the hypermuliplet is of course unsatisfactory. 

The other way to get rid of this term is by taking $v^M$ to be a null vector and imposing the Weyl projection 
\ben
\Gamma_M \eps v^M &=& 0\label{W2}
\een
and again this commutator term will vanish. The advantage of the null reduction is clearly that we can keep the full tensor multiplet structure with the five scalar fields intact.

\section{The Euclidean M5 brane}
So far we have discussed only the Lorentzian M5 brane. But if we eventually would like to study the M5 brane on say $S^6$, then we will need to understand what the Euclidean M5 brane really means in terms of its tensor multiplet structure and its supersymmetry. So here we will clarify this point. First we begin with what is familiar to us though, namely the Lorentzian tensor multiplet and then we seek a way to modify this so that we can allow a Euclidean signature. 

\subsection{The Lorentzian $(2,0)$ and $(0,2)$ tensor multiplets}
We begin with Lorentzian $SO(1,5) \times SO(5) \subset SO(1,10)$ where we have the Dirac conjugate $\bar\eps = \eps^\dag \Gamma^0$ and the Majorana condition $\bar\eps = \eps^T C$ that in terms of Weyl components reads $\eps^{\mp\dag} \Gamma = \eps^{\mp T} C$ and hence is compatible with Weyl projection $\eps^+ = 0$. We then have the chiral $(2,0)$ tensor mutliplet
\bea
\delta \phi^{+A} &=& i \bar\eps \Gamma^A \psi^+\cr
\delta B^+_{MN} &=& i \bar\eps \Gamma_{MN} \psi^+\cr
\delta \psi^+ &=& \frac{1}{12} \Gamma^{MNP} \eps H^+_{MNP} + \Gamma^M \Gamma^A \eps D_M \phi^{+A} - 4 \Gamma^A \eta \phi^{+A}
\eea
We may also consider the anti-chiral $(0,2)$ tensor multiplet
\bea
\delta \phi^{-A} &=& i \bar\eps \Gamma^A \psi^-\cr
\delta B^-_{MN} &=& i \bar\eps \Gamma_{MN} \psi^-\cr
\delta \psi^- &=& \frac{1}{12} \Gamma^{MNP} \eps H^-_{MNP} + \Gamma^M \Gamma^A \eps D_M \phi^{-A} - 4 \Gamma^A \eta \phi^{-A}
\eea
and if we put them together we can write a Lagrangian
\bea
\L_{(2,0)+(0,2)} &=& - \frac{1}{24} H_{MNP}^2 + \L^+ + \L^-\cr
\L^{\pm} &=& - \frac{1}{2} (D_M \phi^{\pm A})^2 - \frac{1}{2} \mu^{AB} \phi^{\pm A} \phi^{\pm B}\cr
&& +  \frac{i}{2} \bar\psi^{\pm} \Gamma^M D_M \psi^{\pm} + \frac{i}{8} \bar\psi^{\pm} \Gamma^{MNP} \Gamma^A \psi^{\pm} T^{\mp A}_{MNP}
\eea
that is invariant under both the $(2,0)$ and the $(0,2)$ superconformal symmetries where the corresponding supersymmetry parameters satisfy
\bea
D_M \eps^\mp &=& \Gamma_M \eta^\pm - \frac{1}{8} \Gamma^A \Gamma^{RST} \Gamma_M \eps^\mp T^{\mp A}_{RST}
\eea
These Killing spinor equations are compatible with the Majorana conditions $\eps^{\mp \dag} \Gamma^0 = \eps^{\mp T} C$ only if we require that
\bea
(\eta^{\pm})^\dag \Gamma^0 &=& (\eta^{\pm})^T C\cr
(T^{\mp A}_{RST})^* &=& T^{\mp A}_{RST}
\eea
To see that we use $(\Gamma^M)^{\dag} = \Gamma^0 \Gamma^M \Gamma^0$.

\subsection{The Euclidean $(2,2)$ tensor multiplet} 
We change to Euclidean signature $SO(6) \times SO(5) \subset SO(6,5)$ by defining the Dirac conjugate as $\bar\eps = \eps^\dag \Gamma$. We impose the 11d Majorana condition $\bar\eps = \eps^T C$ with that new Dirac conjugate. In terms of Weyl components, this reads $(\eps^{\pm})^\dag \Gamma = (\eps^{\mp})^T C$ and we can not impose the 6d Weyl condition. We have the Euclidean nonchiral $(2,2)$ multiplet
\bea
\delta \phi^{\pm A} &=& i \eps^{\pm\dag} \Gamma \Gamma^A \psi^{\pm}\cr
\delta B_{MN} &=& \eps^{\dag} \Gamma \Gamma_{MN} \psi\cr
\delta \psi^\pm &=& \frac{i}{12} \Gamma^{MNP} \eps^\mp H_{MNP} + \Gamma^M \Gamma^A \eps^\mp D_M \phi^{\pm A} - 4 \Gamma^A \eta^\pm \phi^{\pm A}
\eea
where we have removed a factor of $i$ from the variation $\delta B^\pm_{MN}$ to make the variation hermitian by using the Majorana condition. We also multiplied $H_{MNP}$ by a factor of $i$ in $\delta \psi$ to make the variation compatible with the Majorana condition with $H_{MNP}$ real. Because of this $i$, there is a change of sign in the kinetic term for the tensor field and the Lagrangian is
\bea
\L_{(2,2)} &=& \frac{1}{24} H_{MNP}^2 + \L^+ + \L^-
\eea
where the matter part looks identical with that of the Lorentzian $(2,0)+(0,2)$ theory if we write the Dirac conjugates as $\psi^T C$. But if we use the new Majorana condition then it will look like 
\bea
\L^{\pm} &=& - \frac{1}{2} (D_M \phi^{\pm A})^2 - \frac{1}{2} \mu^{AB} \phi^{\pm A} \phi^{\pm B}\cr
&& + \frac{i}{2} \psi^{\mp\dag} \Gamma \Gamma^M D_M \psi^{\pm} - \frac{1}{8} \psi^{\mp\dag} \Gamma \Gamma^{MNP} \Gamma^A \psi^{\pm} T^{\mp A}_{MNP}
\eea
where we also multiplied $T^A_{MNP}$ with a factor of $i$, which is in line with having the same factor of $i$ multiplying $H_{MNP}$. We may notice that the chiral parts $H^{\pm}_{MNP}$ will be complex fields, but the sum, $H_{MNP} = H_{MNP}^+ + H_{MNP}^-$ will be real. This observation may be used for holomorphic factorization of the partition function in Euclidean signature. We get back to the $(2,0)$ tensor multiplet by replacing $\psi^{-\dag} \Gamma$ with $\psi^{+T} C$. Once we have done that replacement, we drop the 11d Majorana condition and impose the Weyl projection $\psi^- = 0$. Then $\L^+$ will become identical with $\L_{(2,0)}$ (although we are now in signature $SO(6,5)$). We can do the corresponding replacements for the $(0,2)$ theory. These two supersymmetries do not mix once we formulate the theory in terms of $\psi^T C$. The supersymmetry parameters satisfy
\bea
D_M \eps^{\mp} &=& \Gamma_M \eta^{\pm} - \frac{i}{8} \Gamma^A \Gamma^{RST} \Gamma_M \eps^{\mp} T^{\pm A}_{RST}
\eea 
where consistency with the Majorana condition implies that 
\bea
\eta^{\pm \dag} \Gamma &=& - \eta^{\pm T} C\cr
(T^{\pm A}_{MNP})^{\dag} &=& T^{\pm A}_{MNP}
\eea

\section{The Majorana condition in various dimensions}
The 11d Majorana condition is
\bea
\bar\psi &=& \psi^T C
\eea
where we define $\bar\psi = \psi^{\dag} \Gamma^t$. We will represent the 11d gamma matrices as
\bea
\Gamma^t &=& i (\sigma^2)^A{}_B \delta^{\alpha}_{\beta} \delta^{\dot\alpha}_{\dot\beta}\cr
\Gamma^m &=& (\sigma^1)^A{}_B (\gamma^m)^{\alpha}{}_{\beta} \delta^{\dot\alpha}_{\dot\beta}\cr
\Gamma^A &=& (\sigma^3)^A{}_B \delta^{\alpha}_{\beta} (\tau^A)^{\dot\alpha}{}_{\dot\beta}
\eea
The charge conjugation matrix is 
\bea
C &=& \eps_{AB} C_{\alpha\beta} C_{\dot\alpha\dot\beta}
\eea
Hence the 11d Majorana condition is
\bea
(\psi^{A\alpha\dot\alpha})^* i (\sigma^2)^A{}_B &=& \psi^{B\beta\dot\beta} \eps_{BA} C_{\beta\alpha} C_{\dot\beta\dot\alpha}
\eea
The 6d chirality matrix is 
\bea
\Gamma &=& (\sigma^3)^A{}_B \delta^{\alpha}_{\beta} \delta^{\dot\alpha}_{\dot\beta}
\eea
So if we define $\eps_{+-} = 1$, then we find 
\bea
(\psi^{+\alpha\dot\alpha})^* &=& C_{\alpha\beta} C_{\dot\alpha\dot\beta}\psi^{+\beta\dot\beta}\cr
(\eps^{-\alpha\dot\alpha})^* &=& C_{\alpha\beta} C_{\dot\alpha\dot\beta}\eps^{-\beta\dot\beta}
\eea
If we reduce to 5d then we have the spinor zero modes that satisfy the above Majorana condition, but the chirality has lost its significance so we choose to not display it when we work in 5d language, so instead of writing $\psi^{+\alpha\dot\alpha}$, we will just write $\psi^{\alpha\dot\alpha}$ when this is a 5d spinor. 

From 
\bea
D_M \eps &=& \Gamma_M \eta
\eea
we get
\bea
D_M \eps^{\dag} \Gamma^t &=& - \eta^{\dag} \Gamma^t\Gamma_M\cr
D_M \eps^T C &=& - \eta^T C \Gamma_M 
\eea
Applying the Majorana condition on the left-hand side of the first equation, we get
\bea
D_M \eps^T C &=& - \eta^{\dag} \Gamma^t \Gamma_M
\eea
and by identifying this with the right hand side of the second equation, we conclude that 
\bea
\eta^{\dag} \Gamma^t &=& \eta^T C
\eea

\section{Metric and Kahler form on $\mb{C}P^2$}\label{Pope}
Here we follow \cite{Gibbons:1978zy}, \cite{Kim:2012tr} and obtain the explicit form of the metric and of the Kahler form on $\mb{C}P^2$. We begin by defining $S^5$ as a sphere that is embedded in $\mb{C}^3$ 
\bea
r^2 &=& |Z^0|^2 + |Z^1|^2 + |Z^2|^2
\eea
with the ambient flat space metric
\bea
ds^2 &=& |dZ^0|^2 + |dZ^1|^2 + |dZ^2|^2
\eea
We define inhomogeneous coordinates
\bea
\zeta^1 &=& \frac{Z^1}{Z^0}\cr
\zeta^2 &=& \frac{Z^2}{Z^0}
\eea
and put
\bea
Z^0 &=& \rho e^{i y}
\eea
where 
\bea
\rho^2 &=& \frac{r^2}{1+\sum_{a=1,2}|\zeta^a|^2}\cr
\eea
and  
\bea
y &\sim & y + 2\pi
\eea
We then get the metric on $S^5$ as
\bea
ds^2 &=& r^2\(\(dy + V\)^2 + \frac{d\zeta^a d\bar\zeta^a}{1+\sum_a |\zeta^a|^2} - \frac{\zeta^a \bar\zeta^b d\bar\zeta^a d\zeta^b}{\(1+\sum_a|\zeta^a|^2\)^2}\)
\eea
where
\bea
V &=& \frac{i}{2\(1+\sum_a |\zeta^a|^2\)} \(\zeta^a d\bar\zeta^a - \bar\zeta^a d\zeta^a\)
\eea
If we parametrize
\bea
\zeta^1 &=& f(\chi,\psi) \cos \frac{\theta}{2} e^{\frac{i\varphi}{2}}\cr
\zeta^2 &=& f(\chi,\psi) \sin \frac{\theta}{2} e^{-\frac{i\varphi}{2}}
\eea
where
\bea
f(\chi,\psi) &=& \tan \chi e^{\frac{i\psi}{2}}
\eea
then we get
\bea
ds^2 &=& r^2 \(dy + V\)^2 + ds_{\mb{CP}^2}^2
\eea
where
\bea
V &=& \frac{1}{2} \sin^2 \chi \sigma_3\cr
ds^2_{\mb{CP}^2} &=& r^2 \(d\chi^2 + \frac{1}{4} \sin^2 \chi \(\sigma_1^2 + \sigma_2^2 + \cos^2\chi \sigma_3^2\)\)
\eea
and  
\bea
\sigma_1 &=& \sin \theta \cos \psi d\varphi - \sin \psi d\theta\cr
\sigma_2 &=& \sin \theta \sin \psi d\varphi + \cos \psi d\theta\cr
\sigma_3 &=& d\psi + \cos \theta d\varphi
\eea
for which we find that 
\bea
d\sigma_3 &=& \sigma_1 \wedge \sigma_2
\eea
and cyclically related relations. We define $\tan \chi \geq 0$ so that $\chi \in [0,\pi/2]$ and we make the identification
\bea
\psi &\sim & \psi + 4\pi
\eea
We define the vielbein
\bea
e^4 &=& r d\chi\cr
e^1 &=& \frac{r}{2} \sin \chi \sigma_1\cr
e^2 &=& \frac{r}{2} \cos \chi \sigma_2\cr
e^3 &=& \frac{r}{2} \sin \chi \cos \chi \sigma_3
\eea
We then find that
\bea
F = dV = \frac{2}{r^2} J
\eea
where
\bea
J &=& e^4 \wedge e^3 + e^1 \wedge e^2
\eea
is the Kahler form.

\section{The vielbein components in lightcone coordinates}
In lightcone coordinates on $\mb{R} \times S^5$, the vielbein has the components
\bea
\(\begin{matrix}
e^{\h{+}}{}_+ & e^{\h{+}}{}_- & e^{\h{+}}{}_i\\
e^{\h{-}}{}_+ & e^{\h{-}}{}_- & e^{\h{-}}{}_i\\
e^{\h{i}}{}_+ & e^{\h{i}}{}_- & e^{\h{i}}{}_i
\end{matrix}\) &=& \(\begin{matrix}
1 & 0 & \frac{r}{\sqrt{2}} \kappa_i\\
0 & 1 & - \frac{r}{\sqrt{2}} \kappa_i\\
0 & 0 & E^{\h{i}}{}_i
\end{matrix}\)
\eea
and its inverse is
\bea
\(\begin{matrix}
e^+{}_{\h{+}} & e^+{}_{\h{-}} & e^+{}_{\h{i}}\\
e^-{}_{\h{+}} & e^-{}_{\h{-}} & e^-{}_{\h{i}}\\
e^i{}_{\h{+}} & e^i{}_{\h{-}} & e^i{}_{\h{i}}
\end{matrix}\) &=& \(\begin{matrix}
1 & 0 & - \frac{r}{\sqrt{2}} \kappa_{\h{i}}\\
0 & 1 & \frac{r}{\sqrt{2}} \kappa_{\h{i}}\\
0 & 0 & E^i{}_{\h{i}}
\end{matrix}\)
\eea
The metric is 
\bea
ds^2 &=& - 2 e^{\h{+}} e^{\h{-}} + e^{\h{i}} e^{\h{i}}
\eea
and $E^{\h{i}}$ denotes the vielbein on $\mb{C}P^2$. Since $\kappa_i$ is a Killing vector, we have the important identity 
\bea
\kappa^i w_{ij} &=& 0
\eea
where $w_{ij}$ is the Kahler form. Here $\kappa$ was denoted as $V$ and $w = d\kappa$ was denoted as $J$ in appendix \ref{Pope}.


\begin{thebibliography}{999}




\bibitem{Kim:2012tr}
H.~C.~Kim and K.~Lee,
``Supersymmetric M5 Brane Theories on R x CP2,''
JHEP \textbf{07}, 072 (2013)
[arXiv:1210.0853 [hep-th]].





\bibitem{Hosomichi:2012ek}
K.~Hosomichi, R.~K.~Seong and S.~Terashima,
``Supersymmetric Gauge Theories on the Five-Sphere,''
Nucl. Phys. B \textbf{865}, 376-396 (2012)
[arXiv:1203.0371 [hep-th]].


\bibitem{Kim:2012ava}
H.~C.~Kim and S.~Kim,
``M5-branes from gauge theories on the 5-sphere,''
JHEP \textbf{05} (2013), 144
[arXiv:1206.6339 [hep-th]].






\bibitem{Lambert:2010wm}
N.~Lambert and C.~Papageorgakis,
``Nonabelian (2,0) Tensor Multiplets and 3-algebras,''
JHEP \textbf{08} (2010), 083
[arXiv:1007.2982 [hep-th]].


\bibitem{Gustavsson:2020ugb}
A.~Gustavsson,
``A nonabelian M5 brane Lagrangian in a supergravity background,''
JHEP \textbf{10} (2020), 001
[arXiv:2006.07557 [hep-th]].



\bibitem{Lambert:2020scy}
N.~Lambert and T.~Orchard,
``Null reductions of the M5-brane,''
JHEP \textbf{12} (2020), 037
[arXiv:2005.14331 [hep-th]].


\bibitem{Gustavsson:2018rcc}
A.~Gustavsson,
``The non-Abelian tensor multiplet,''
JHEP \textbf{07}, 084 (2018)
[arXiv:1804.04035 [hep-th]].



\bibitem{Gibbons:1978zy}
G.~W.~Gibbons and C.~N.~Pope,
``CP**2 AS A GRAVITATIONAL INSTANTON,''
Commun. Math. Phys. \textbf{61} (1978), 239

\end{thebibliography}
\end{document}